\documentclass[%
reprint,
superscriptaddress,
amsmath,
amssymb,
aps,
pra,
floatfix,
]{revtex4-2}

\usepackage{graphicx}
\usepackage{dcolumn}
\usepackage{bm}
\usepackage{hyperref}
\usepackage{placeins}
\usepackage{physics}
\usepackage[english]{babel}
\usepackage[caption=false]{subfig}

\usepackage{xcolor}
\usepackage{dsfont}
\usepackage{amsthm}
\usepackage{braket}
\usepackage{optidef}
\usepackage{algorithm2e}
\usepackage[toc,page]{appendix}
\usepackage{wasysym}
\usepackage{tikz}

\makeatletter
\let\ORIbbl@fixname\bbl@fixname
\def\bbl@fixname#1{%
  \@ifundefined{languagealias@\expandafter\string#1}
    {\ORIbbl@fixname#1}
    {\edef\languagename{\@nameuse{languagealias@#1}}}%
}
\newcommand{\definelanguagealias}[2]{%
  \@namedef{languagealias@#1}{#2}%
}
\makeatother

\definelanguagealias{en}{english}
\definelanguagealias{EN}{english}
\definelanguagealias{en-GB}{english}

\makeatletter
\newcount\my@repeat@count
\newcommand{\rep}[2]{%
  \begingroup
  \my@repeat@count=\z@
  \@whilenum\my@repeat@count<#1\do{#2\advance\my@repeat@count\@ne}%
  \endgroup
}
\makeatother

\definecolor{LB}{RGB}{134,41,198}

\begin{document}

\preprint{APS/123-QED}

\title{Quantum spatial search with multiple excitations}

\author{Dylan Lewis}
 \email{dylan.lewis.19@ucl.ac.uk}
\affiliation{Department of Physics and Astronomy, University College London, London WC1E 6BT, United Kingdom}

\author{Leonardo Banchi}
\affiliation{Department of Physics and Astronomy, University of Florence, 
via G. Sansone 1, I-50019 Sesto Fiorentino (FI), Italy}
\affiliation{INFN Sezione di Firenze,  via G. Sansone 1,  I-50019 Sesto Fiorentino (FI), Italy}

\author{Sougato Bose}
\affiliation{Department of Physics and Astronomy, University College London, London WC1E 6BT, United Kingdom}

\begin{abstract}
Spatial search is the problem of finding a marked vertex in a graph. 
A continuous-time quantum walk in the single-excitation subspace of an $n$ spin system solves the problem of spatial search by finding the marked vertex in $O(\sqrt{n})$ time. Here, we investigate a natural extension of the spatial search problem, marking multiple vertices of a graph, which are still marked with local fields. We prove that a continuous-time quantum walk in the $k$-excitation subspace of $n$ spins can determine the binary string of $k$ marked vertices with an asymptotic fidelity in time $O(\sqrt{n})$, despite the size of the state space growing as $O(n^k)$. Numerically, we show that this algorithm can be implemented with interactions that decay as $1/r^\alpha$, where $r$ is the distance between spins, and an $\alpha$ that is readily available in current ion trap systems.
\end{abstract}

\maketitle

\section{\label{sec:intro}Introduction}
An unstructured search problem of finding a particular marked element in a database of $n$ elements takes a classical computer $O(n)$ time. Grover's algorithm famously showed this problem can be solved in time $O(\sqrt{n})$ by a quantum computer~\cite{grover_fast_1996}, which is optimal~\cite{boyer_tight_1998}. Farhi and Gutmann proposed an analogue version of Grover search, which solved the search problem with a continuous-time quantum walk on the complete graph of states~\cite{Farhi1998}. Childs and Goldstone showed that this was equivalent to solving the spatial search problem~\cite{Childs2004}, where a local field marks a particular spin in a spin graph, and the problem is finding this marked spin. 
The spin graph can be defined in the single-excitation subspace of $n$ spins with spin 1/2, i.e.~qubits~\cite{childs_spatial_2004}. 
The edges in the graph represent interactions between spins, the hopping of the continuous-time quantum walk. Many spin graphs have been found to permit optimal spatial search: the hypercube graph, $d$-dimensional periodic lattices with $d>4$~\cite{Childs2004}, sufficiently well-connected spin graphs~\cite{chakraborty_optimality_2020},
and various other high-dimensional graphs~\cite{novo_systematic_2015,chakraborty_spatial_2016,chakraborty_optimal_2017,novo_environment-assisted_2018, wong_quantum_2018,osada_continuous-time_2020, sato_scaling_2020,malmi_spatial_2022}. 
 Being \emph{optimal} means the marked spin can be found in time $O(\sqrt{n})$.
Recently, optimal spatial search was found for one-dimensional spin chains with long-range interactions~\cite{lewis_optimal_2021}, which presented the possibility for an experimental realisation of solving the spatial search problem in optimal time~\cite{lewis_ion_2023}. 

The spatial search problem is an unstructured search of a database containing binary strings $j$, which represent the spin configurations. In the single-excitation subspace, 
$\ket{j} = \ket{0}^{\otimes(j{-}1)}{\otimes}\ket1{\otimes}\ket{0}^{\otimes(n{-}j)}= \ket{0...010...0}$ and the database contains $n$ entries. 
A more useful and unexplored problem in spatial search is for a database containing binary strings with $k>1$ excitations, namely ``1s'', such as $\ket{j} = \ket{0100100}$ for $k=2$. A clear advantage of this database is that it encodes more binary strings into the same number of spins. Considering the possibility of experimental implementation with long-range interactions in ion traps, search with additional spin excitations offers more practical use cases, such as sensing magnetic defects in spin lattices, or a form of fast quantum QR code reader, offering an alternative route to explore quantum advantage in pattern detection than that already discussed for optical systems~\cite{banchi2020quantum}. The problem now is whether a continuous-time quantum walk algorithm can find a specific binary string in the database of elements marked only by $k$ local fields at the positions of the 1s. The subspace size now scales as $O(n^k)$. Remarkably, we will show that the marked element can still be found in time $O(\sqrt{n})$ for $k>1$. The reason for this is that the problem is now a structured search and the optimality of Grover search no longer applies. 

In the single-excitation subspace, the spatial search problem 
is encoded in the marking Hamiltonian $H_\textrm{mark}$, which marks the local site $w$ with a Hamiltonian term proportional to $|w\rangle\langle w|$, where $w$ is a binary string with a single 1. The walk Hamiltonian $H$ introduces the hopping between states, giving the edges of the state graph. $H$ is also time-independent and therefore generates the evolution of a continuous-time quantum walk on the state graph. The spatial search algorithm starts by initialising the graph in the state $\ket{s}$, which is an equal superposition of all states in the single-excitation subspace. The overall Hamiltonian is the sum of the marking and walk Hamiltonians, $H_\textrm{search} = \gamma H + H_\textrm{mark}$, where $\gamma$ is the hopping rate. The $\gamma$ is chosen such that, if $H$ permits optimal spatial search, after the state $|s\rangle$ is evolved for a time $T = O(\sqrt{n})$, we find the state of the system is $|w\rangle$ with a high fidelity $F$. The fidelity is $F(t) = |\langle w| e^{-iH_\textrm{search}t}|s\rangle|^2$. 

To design a quantum algorithm for our problem, we invoke a new marking Hamiltonian $H_\textrm{mark}$ that has local fields at $k$ sites. The same walk Hamiltonian of the graph now takes place in the $k$-excitation subspace, and the initial state is an equal superposition of all states in the subspace. The algorithm is otherwise equivalent to the spatial search case.

\section{\label{sec:preliminaries}Hamiltonian}
We first recall a few basic facts. 
A spin simply refers to a two-level quantum system, also called qubit. 
It has two basis states, which in the computational basis, we write as $|0\rangle$ and $|1\rangle$. We define the term \textit{excitation} as a spin being in state 1, and the term \textit{ground state} corresponds to the spin being in state 0. The basis states of $n$ spins can be represented as binary strings. Binary strings are elements of the set of all combinations of $0$ and $1$ of length $n$ -- we define this set $\mathcal{V} = \{0,1\}^n$. For spatial search in $\mathcal V$, the marking operator is $n$-local, namely it contains the tensor product of $n$ Pauli operators, making its simulation quite challenging. On the other hand, it is desirable to focus on spin-local marking operators, with a single non-trivial Pauli operator -- for example, $\sigma^z_i$, which is a Pauli $z$ operator acting on the $i$th spin. We consider spin excitation subspaces, where the number of excitations is fixed such that the number of 1s in the binary string is $k$, $\mathcal{V}_k = \set{ b \in \mathcal{V} | \sum_i^n b_i = k} $, where $b_i$ is the value of the $i$th bit of $b$.
The spin-local marking operator is equal to the standard case of spatial search with a single marked site for $\mathcal V_1$.

For spatial search in $\mathcal V_1$, the number of spins required is equal to the number of states, since the set of binary strings has size $|\mathcal{V}_1|=n$. The basis states are a single excitation at any spin. This is a spatially inefficient coding. Spatially-encoded quantum search in higher-excitation subspaces would encode more binary strings into the same number of spins. For $k$ marked sites, we can encode binary strings composed of a $k$ number of 1s and $n-k$ number of 0s. This is the $k$-excitation subspace and gives a binary string state space of size $N_k = |\mathcal{V}_k| = \binom{n}{k}$, and with $N = \sum_k N_k = 2^n$. We introduce the marking Hamiltonian
\begin{align}
    \label{eq:mark_Hamiltonian}
    H_\mathrm{mark} = -\frac{1}{2}\sum_{i \in \mathcal{M}} \sigma^z_i, 
\end{align}
where $\mathcal{M}$ is the set of marked sites, with $k = |\mathcal{M}|$. Physically, this can be encoded by a local field at the marked sites. In Section~\ref{sec:experimental}, we discuss the experimental implementation in more detail.
The location of the marked sites identify a single binary string with a 1 at $m$ for all $m \in \mathcal{M}$ and 0 for all other bits in the string. We label this marked string $w$, and the state corresponding to an excitation at each $m$ is $|w\rangle$. The binary strings define vertices of a graph, such that each vertex is labelled by a binary string. The set of vertices is $\mathcal{V}_k$. The edges of the graph correspond to the hopping of spin excitations due to an XY model Hamiltonian, which we call the walk Hamiltonian,
\begin{align}
    \label{eq:walk_Hamiltonian_spin}
    H = \frac{1}{4} \sum_{i\ne j} J_{ij} \left(\sigma^x_i\sigma^x_j + \sigma^y_i\sigma^y_j \right),
\end{align}
where $J_{ij}$ defines the strength of the interaction between spin $i$ and $j$. This Hamiltonian commutes with $\sum_i^n\sigma^z_i$, which means the total spin excitation number is conserved. Since $H_\mathrm{mark}$ also commutes with this term, the evolution of both Hamiltonians is confined to the excitation subspace of the initial state. It can be made more clear that the Hamiltonian $H$ can be represented as edges of an undirected graph $G_k$ with vertices in $\mathcal{V}_k$ by viewing the Hamiltonian in the $k$-excitation subspace, 
\begin{equation}
    \label{eq:walk_Hamiltonian}
    H = \sum_{ (a,b) \in \mathcal{E}_k} J^\prime_{a b} \left(|a\rangle\langle b| + h.c.\right) ,
\end{equation}
where $\mathcal{E}_k = \set{ (a,b) | a,b \in \mathcal{V}_k, d(a,b) = 2 }$ is the set of pairs of vertices with Hamming distance $d(a,b) = 2$, where Hamming distance measures the difference between two binary strings $a$ and $b$. Hamming distance 2 means that only two bits have been flipped, which in our case can only be caused by a single hop of an excitation. The set $\mathcal{E}_k$ gives the edges of the graph $G_k$. The summation of Eq.~\eqref{eq:walk_Hamiltonian} is now over binary strings $a$ and $b$ rather than spin labels as in Eq.~\eqref{eq:mark_Hamiltonian} and the strength of the interaction is now $J^\prime_{a b}$ rather than $J_{ij}$. The mapping between these interactions is given by 
\begin{align}
    \label{eq:J_ab_J_ij_mapping}
    J^\prime_{a b} = \sum_{i,j} J_{ij} (a_i - b_i)(b_j-a_j).
\end{align}
The expression can be understood by the term $(a_i - b_i)(b_j-a_j)$ only being non-zero if the bits at the $i$th and $j$th site have both flipped. Since $a$ and $b$ are binary strings that have an edge, the summand in Eq.~\eqref{eq:J_ab_J_ij_mapping} finds the spins $i$ and $j$ that the excitation hops between for the edge $(a,b)$ in the graph. The marked Hamiltonian can also be considered as a summation over the states as binary strings
\begin{align}
    \label{eq:mark_Hamiltonian_strings}
    H_\textrm{mark} = \sum_{a \in \mathcal{V}_k} c_k(a) |a\rangle\langle a|,
\end{align}
where we have defined a function, 
\begin{align}
    c_k(a) =  \left(\sum_{m\in \mathcal{M}} a_m \right) -\frac{k}{2},
\end{align}
that counts the number of excitations in the state $a$ that are on the marked sites of $\mathcal{M}$ and then subtracts the constant $k/2$. In the general case, the graph $G_k$ is weighted by $J_{ab}$ between vertices $a$ and $b$. First, we consider the unweighted case of interactions between all spins with equal strength. 

\begin{figure}
    \centering
    \includegraphics[width=0.85\linewidth]{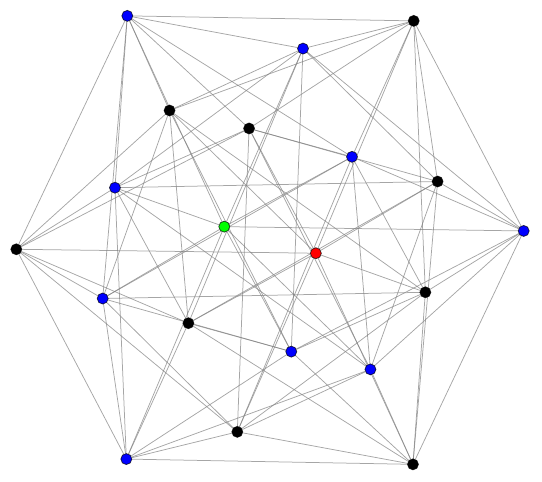}
    \caption{Johnson graph for $J(6,3)$. The vertices represent the states for a system of all-to-all interactions with 6 spins and 3 excitations. The red vertex indicates the state $|w\rangle$; the black vertices are graph distance 1 from $w$; the blue vertices are graph distance 2 from $w$; and the green vertex is graph distance 3 from $w$. Equivalently, the red vertex is the state where all excitations of the spin state are precisely on local fields; the black vertices are the sites where a single excitation of the spin state is not on a local field; the blue vertices are where two excitations of the spin state are not on the local fields; and the green vertices are where three excitations of the spin state are not on the local fields.}
    \label{fig:johnson_6_3}
\end{figure}

The Johnson graph $J(n,k)$ is defined as having vertices that are the $k$-element subsets of $n$ elements and edges that have an intersection of $(k-1)$ elements between two vertices. The vertices of $J(n,k)$ are therefore equivalent to $\mathcal{V}_k$ and the edges are the hopping of a single excitation of $\mathcal{E}_k$. We therefore have $G_k = J(n,k)$ for fully-connected spins. A Johnson graph $J(6,3)$ is given as an example in Fig.~\ref{fig:johnson_6_3}. More examples of Johnson graphs are given in Fig.~\ref{fig:johnson_graphs_Jn2} in App.~\ref{app:spatial_multiple_excitations}. Spatial search in the single-excitation subspace with all-to-all interactions is therefore given by $J(n,1)$, which is the complete graph with $n$ vertices. 

The initial state, $|s_k\rangle = \frac{1}{\sqrt{N_k}}\sum_{a \in \mathcal{V}_k} |a\rangle$, is the equal superposition of the states corresponding to the vertices of $G_k$. Full details for the following expressions are given in App.~\ref{app:spatial_multiple_excitations}. The set of vertices a distance $q$ from the marked vertex $w$ in the graph is $\mathcal{D}_{k,q}^{(w)} = \set{v\in\mathcal{V}_k | d(w,v) = 2q }$, where $d(w,v)$ is the Hamming distance between $w$ and $v$. The cardinality of the sets is $d_{k,q} = \vert \mathcal{D}_{k,q}^{(w)}\vert = \binom{k}{q}\binom{n-k}{q}$. We introduce the state $|d_{k,q}\rangle$ that is an equal superposition of the basis states a distance $q$ from the marked state, i.e. an equal superposition of the basis states corresponding to the elements of the set $\mathcal{D}_{k,q}^{(w)}$.

The evolution of the system with initial state $|s_k\rangle$ can be written in a simplified $(k+1)$-dimensional subspace $\mathcal W_k =\{|w\rangle, |d_{k,1}\rangle, |d_{k,2}\rangle, ..., |d_{k,k}\rangle  \}$. The elements of the search Hamiltonian in the subspace are found using combinatorics, described in App.~\ref{app:combinatorics}. The result is a  matrix 
\begin{align}
    H_{i j} &= \begin{cases}
         (j-1)(n-2j+2) & \textrm{for } i=j,\\
         i\sqrt{(k-i+1)(n-k-i+1)} & \textrm{for } j-i=1,\\
         0 & \textrm{otherwise},
    \end{cases} \label{eq:subspace_hamiltonian}
\end{align}
for $i,j\in\mathcal W_k$, 
which, for the initial state $|s_k\rangle$, gives the exact evolution. We show in App.~\ref{app:spin_formalism} that this Hamiltonian can also be derived with the spin formalism. After defining the global spin operators $\vec{S}_1 = \frac{1}{2}\sum_{i \in \mathcal{M}} \vec{\sigma}$, $\vec{S}_2 = \frac{1}{2}\sum_{i \in \mathcal{M}} \vec{\sigma}$, and $\vec{S} = \vec{S}_1 + \vec{S}_2$, the walk Hamiltonian can be expressed as $H = \vec{S}\cdot\vec{S} - (S^z)^2 - \frac{n}{2} \mathds{1}$ and marked Hamiltonian as $H_\textrm{mark} = S^z_1$. Restricting marked and unmarked spins to their symmetric subspaces, the search Hamiltonian can be expressed in a $(k+1)\times(k+1)$ subspace that is equivalent to Eq.~\eqref{eq:subspace_hamiltonian}. 
In this language, the subspace $\mathcal W_k$ is identified by the possible spin eigenvalues of the global spin operator $S^z_1$.

\section{\label{sec:asymptotic_fidelity}Asymptotic Fidelity}
\begin{figure}[htb]
    \centering
    \includegraphics[scale=0.32]{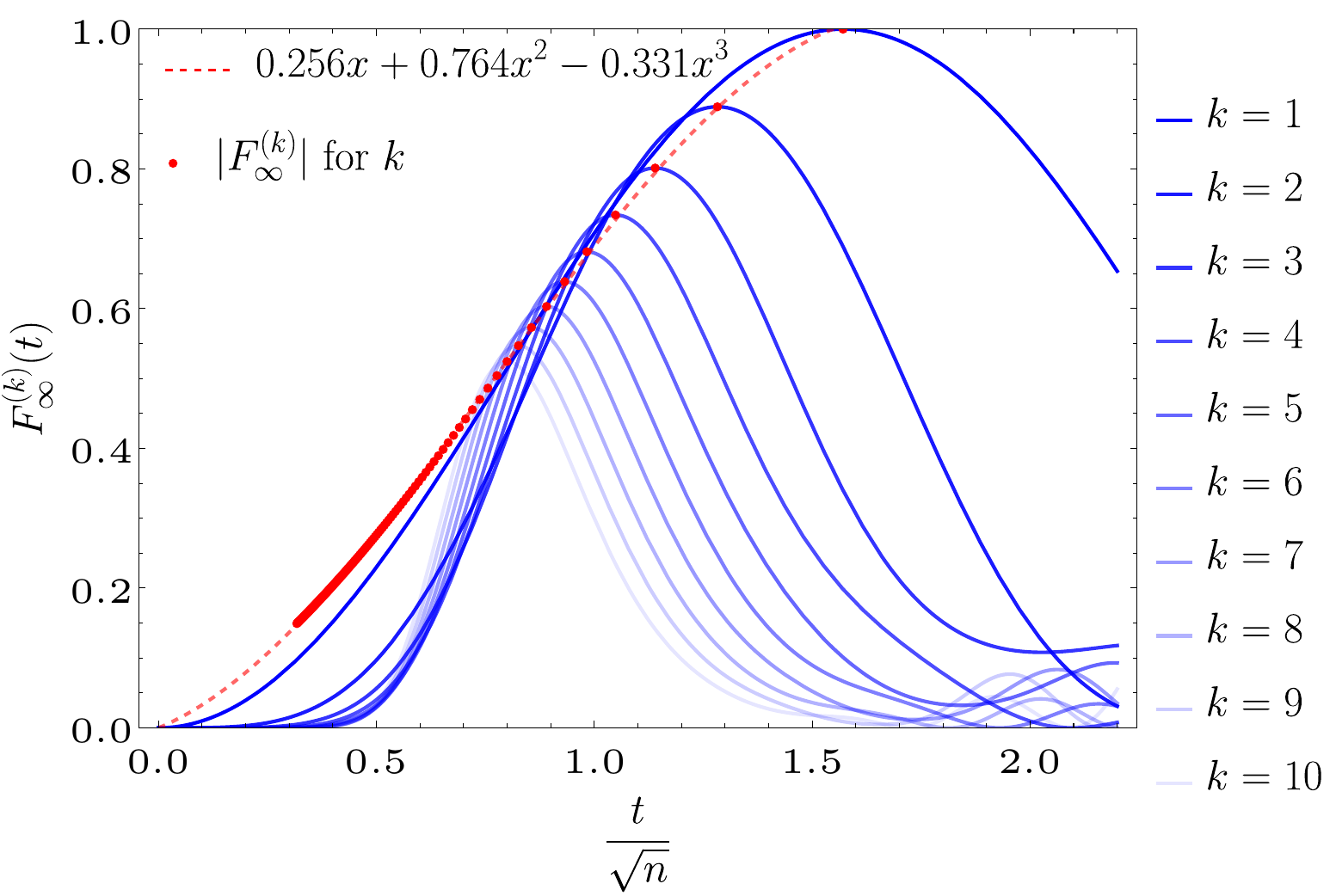}
    \caption{Plot of $F_\infty^{(k)}(t)$ for $k=2$ up to $k=10$. The increasing $k$ is shown with decreasing opacity. The time is rescaled by $1/\sqrt{n}$ such that only $R_k$ is required, which is generated according to Eq.~\eqref{eq:R_k_matrix_elements}. $\gamma = 1/n$ has been used for the asymptotic expressions.}
    \label{fig:asymptotic_fidelities_k}
\end{figure}

\begin{figure}[htb]
    \centering
    \subfloat[Time to reach asymptotic maximum fidelity]{\includegraphics[scale=0.42]{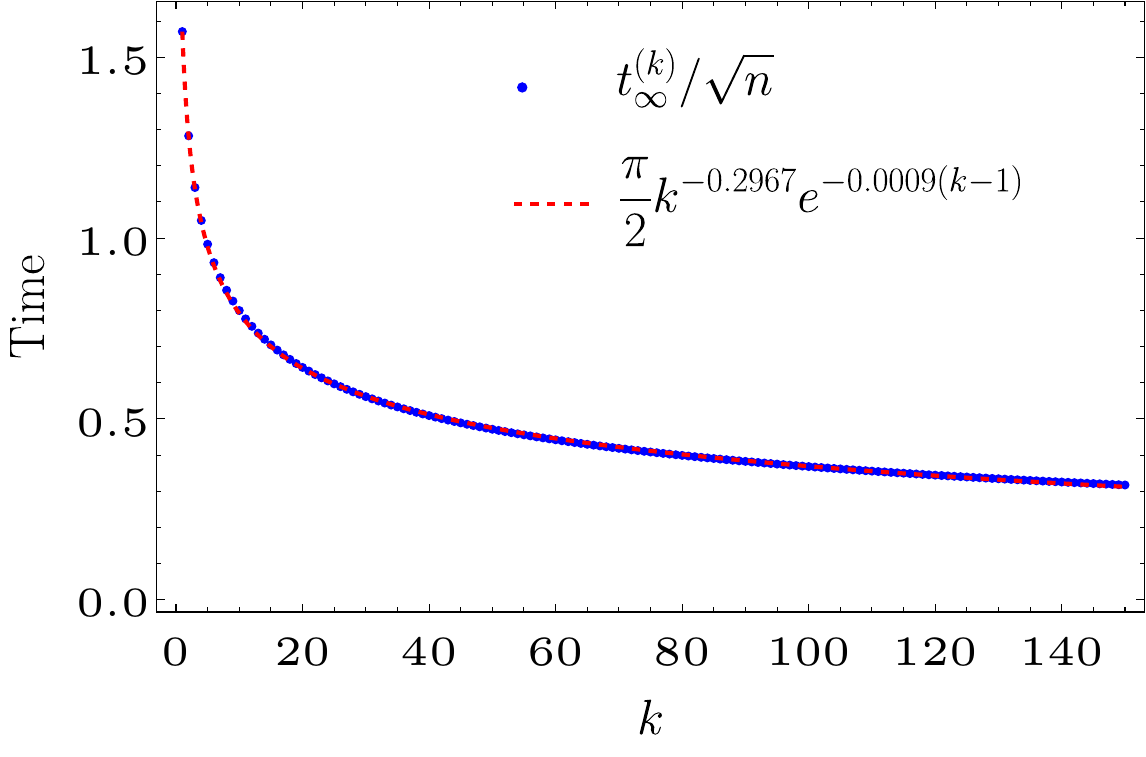}}
    
    \subfloat[Asymptotic maximum fidelity]{\includegraphics[scale=0.42]{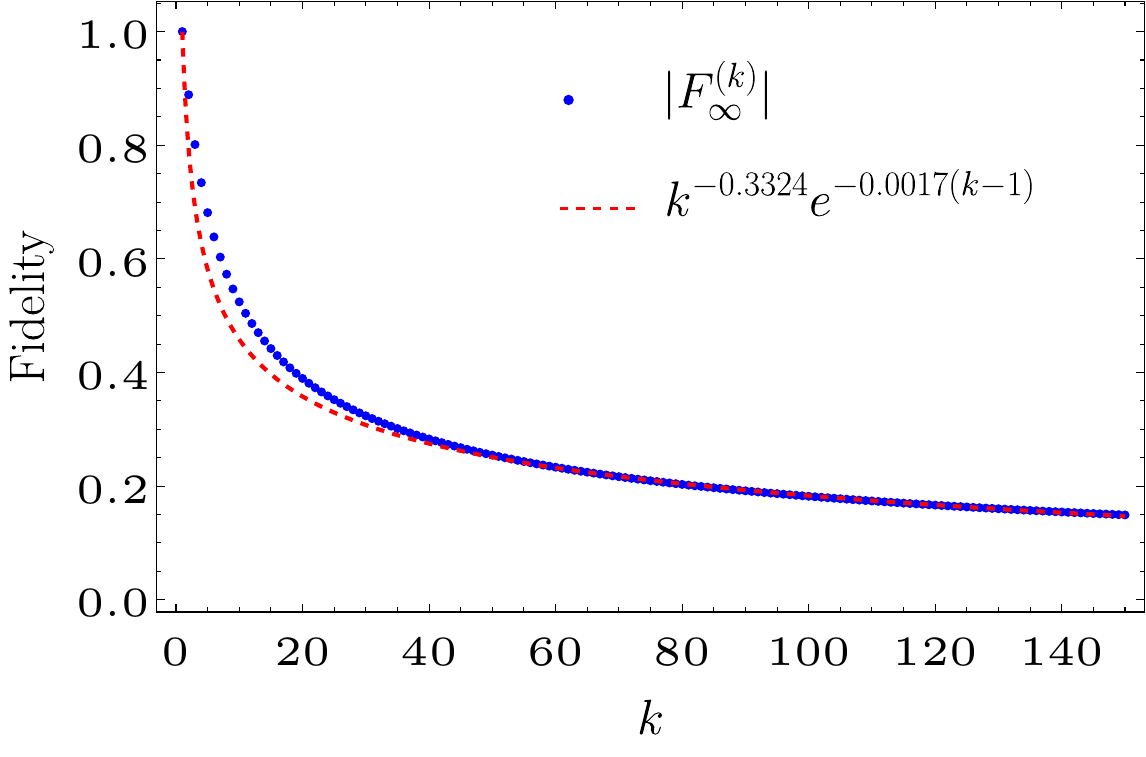}}
    \caption{Plot (a) is the asymptotic time $t^{(k)}_\infty/\sqrt{n}$ to reach the maximum fidelity against $k$, with a simple best fit model. Plot (b) is for the asymptotic maximum fidelity $|F_\infty^{(k)}|$ as a function of $k$ with an simple fit that is appropriate for large $k$, but underestimates the fidelity for low $k$. }
    \label{fig:asymptotic_fidelities_and_times_against_k}
\end{figure}

The fidelity of search with $k$ excitations is defined as 
\begin{align}
    F^{(k)}(t) = \vert \langle w | e^{-i H_\textrm{search}t }| s_k \rangle \vert^2.
\end{align}
We define the asymptotic fidelity $F^{k}_\infty(t)$ as the first order expansion about large $n$ of the fidelity $F^{(k)}(t)$. The maximum fidelity is defined as $|F_\infty^{(k)}| = \max_{t\in \mathbb{R}^+}\{F_\infty^{(k)}(t)\}$. The time to reach the maximum fidelity is $t^{(k)}_\infty$. In the limit of large $n$, the initial state becomes $\lim_{n\rightarrow\infty}|s_k\rangle = |d_{k,k}\rangle$. We can relabel the states $|w\rangle$ and $|d_{k,k}\rangle$ by $|1\rangle$ and $|k+1\rangle$ respectively to correspond to basis vectors of the simplified subspace. Setting $\gamma = \frac{1}{n}$, we find the asymptotic fidelity for subspace $k$ is 
\begin{align}
    F^{(k)}_\infty (t) = \big\vert \langle 1| e^{-i\frac{R_k}{\sqrt{n}}t} |k+1\rangle \big\vert^2,
\end{align}
with the asymptotic evolution matrix $R_k \in \mathbb{R}^{(k+1)\times(k+1)}$, with elements 
\begin{align}
    (R_k)_{ij} = j\sqrt{k-j+1}\delta_{i,j+1} + i \sqrt{k-i+1}\delta_{j,i+1},
    \label{eq:R_k_matrix_elements}
\end{align}
which is surprisingly similar to a Hamiltonian allowing perfect state transfer \cite{albanese2004mirror}, yet 
without the mirror symmetry. 
The maximum asymptotic fidelity for $k=2$ is $|F_\infty^{(2)}| = \frac{8}{9}$, which is derived in App.~\ref{app:two_excitation_subspace}. In App.~\ref{app:three_excitation_subspace}, we show $|F_\infty^{(3)}| \geq 0.702$. Fig.~\ref{fig:asymptotic_fidelities_k} shows the asymptotic fidelity for subspaces $k=1,...,10$. In Fig.~\ref{fig:asymptotic_fidelities_and_times_against_k}, we show how the asymptotic fidelity and the time to reach the asymptotic fidelity scale with $k$. The asymptotic fidelity decreases as the subspace $k$ increases. However, the time to reach the maximum fidelity still scales as $t_\infty^{(k)} = O(\sqrt{n})$, even with the large subspace size of $N_k = O(n^k)$. The best fit curves of Fig.~\ref{fig:asymptotic_fidelities_and_times_against_k} indicate that the fidelity tends to 0 as $k$ approaches $\frac{n}{2}$ asymptotically. 

\section{\label{sec:experimental}Experimental implementation and decaying interactions}
Experimental platforms typically have limited capability to generate all-to-all interactions between spins. While for a low number of spins (up to around 10) in ion traps, it is possible to generate complete graphs of interactions and other complex interaction graphs~\cite{teoh_machine_2020}, as the number of spins increases there is a power law decay in the interaction strength with the distance between ions. The decay in interaction strength is due to the closing of the spectrum of phonon modes that mediate the spin-spin interactions. As the number of spins increases, the number of phonon modes increases, and it becomes increasingly hard to only excite the centre-of-mass phonon mode (which has no decay of interaction strength)~\cite{lewis_ion_2023}. As more phonon modes are excited the interaction strength decreases with the distance as a power law $1/r^\alpha$, where $r$ is the distance between the ions and $\alpha$ is a parameter that characterises the interaction range. In the limit of only the centre-of-mass phonon mode contributing, $\alpha$ is $0$, giving a complete graph of interactions. As more phonon modes are excited, $\alpha$ tends to $3$. For chains of up to 50 ions, $\alpha = 1$ is readily achievable~\cite{friis_observation_2018, monroe_programmable_2021}. 

For 10 ions (spins) the spatial search problem in the single-excitation subspace only searches 10 states. However, in higher excitation subspaces, the number of states for 10 spins can be significantly higher; $k=2$ has $45$ states, and $k=3$ has $120$ states. In general, there are $N_k = \binom{n}{k}$ for $n$ spins with $k$ excitations. Since $\alpha$ increases as the ion chain gets longer (as $n$ increases), a larger number of states is possible with higher $k$ whilst keeping $\alpha$ low enough to maintain effective long-range interactions. The spatial search problem itself becomes more structured as $k$ increases, discussed further in the following Section. Long-range interactions on a 1-dimensional periodic spin chain can be modelled by a coupling strength
\begin{align}
    J_{ij} = \frac{1}{2} \left( \frac{1}{|i-j|^\alpha} + \frac{1}{(n-|i-j|)^\alpha} \right),
    \label{eq:long_range_coupling}
\end{align}
between spins $i$ and $j$, giving a coupling $J_{ij} \sim 1/r^\alpha$ with $r$ being the distance between the spins. In Fig.~\ref{fidelities_for_n_12_against_alpha}, we show numerically that high fidelity spatial search is still possible as $\alpha$ increase for $n=12$ and subspaces $k=1,2,3,4,5$. Fig.~\ref{fig:fidelities_for_low_n_against_alpha} in App.~\ref{app:long_range_interaction_graphs} shows further results for $n=10$ and $n=11$.
\begin{figure}[tb]
    \centering
    \includegraphics[scale=0.42]{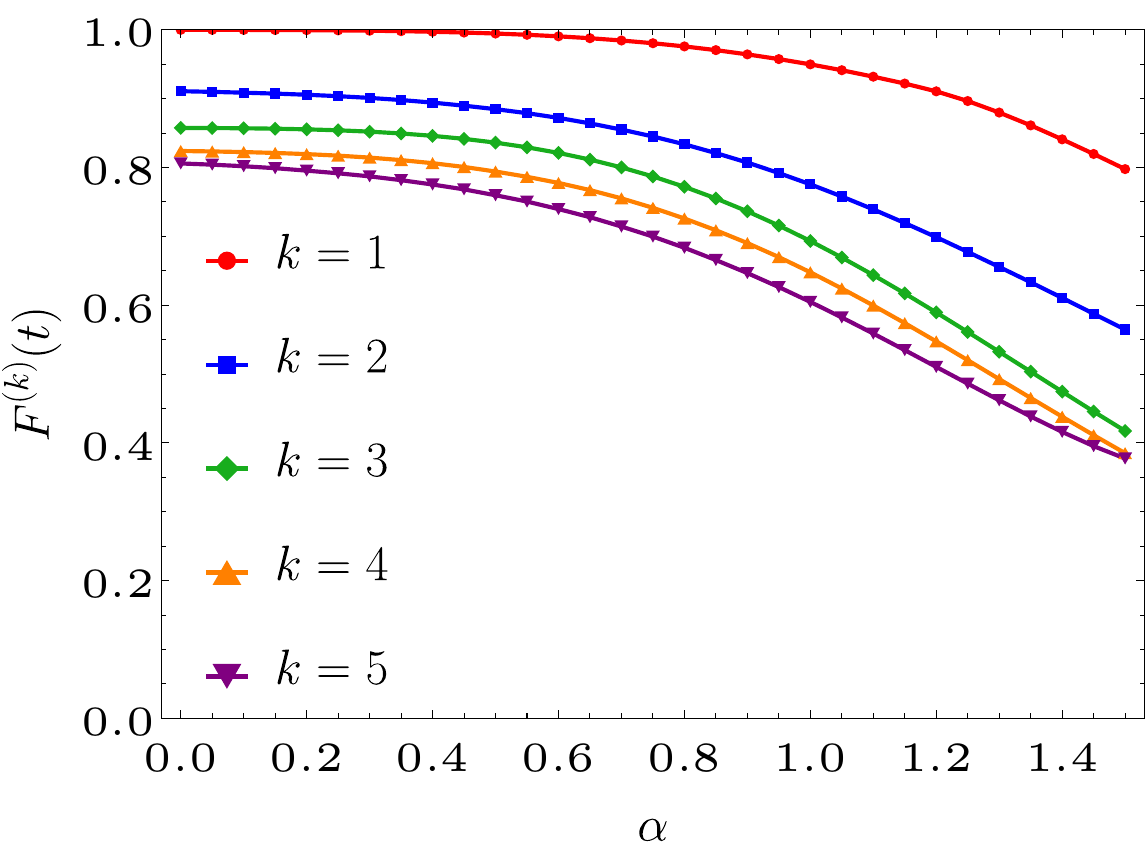}
    \caption{Plot of $F^{(k)}(t)$ with $n=12$ for varying interaction range, where the interactions are defined in Eq.~\eqref{eq:long_range_coupling}, parameterised by $\alpha$. Fidelity for $k=1$ (red), $k=2$ (blue), $k=3$ (green), $k=4$ (orange), and $k=5$ (purple). The hopping rate $\gamma$ is optimised for each $\alpha$ using a line search.}
    \label{fig:fidelities_for_n_12_against_alpha}
\end{figure}

The argument in Ref.~\cite{lewis_optimal_2021} rules out search in time $O(\sqrt{n})$ for $\alpha > 1.5$ when $k=1$ and dimension $d=1$, as the Lieb-Robinson bounds for free-particle long-range interactions~\cite{tran_hierarchy_2020} give an effective light cone for a maximum correlation distance. However, for $k>1$, it is not clear the same bound would apply. The more general Frobenius light cone would preclude spatial search in time $O(\sqrt{n})$ for $\alpha > 2$.

\section{\label{sec:repeated_spatial_search}Advantage over repeated single-excitation spatial search}
\begin{figure*}[htb]
    \centering
    \subfloat[Number of spins $n$]{\includegraphics[scale=0.38]{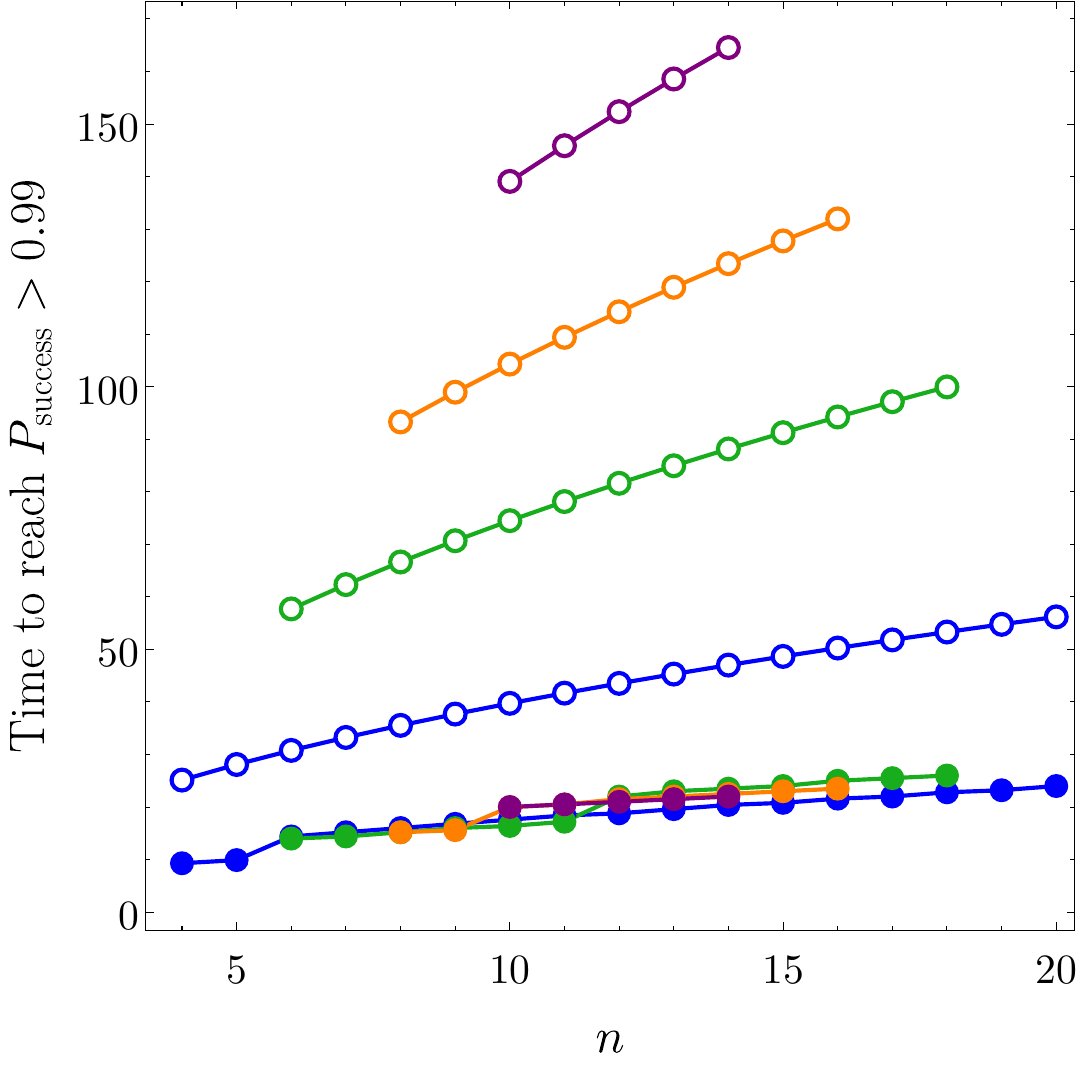}}
    \hspace{2mm}
    \subfloat[Number of states $N_k = \binom{n}{k}$]{\includegraphics[scale=0.38]{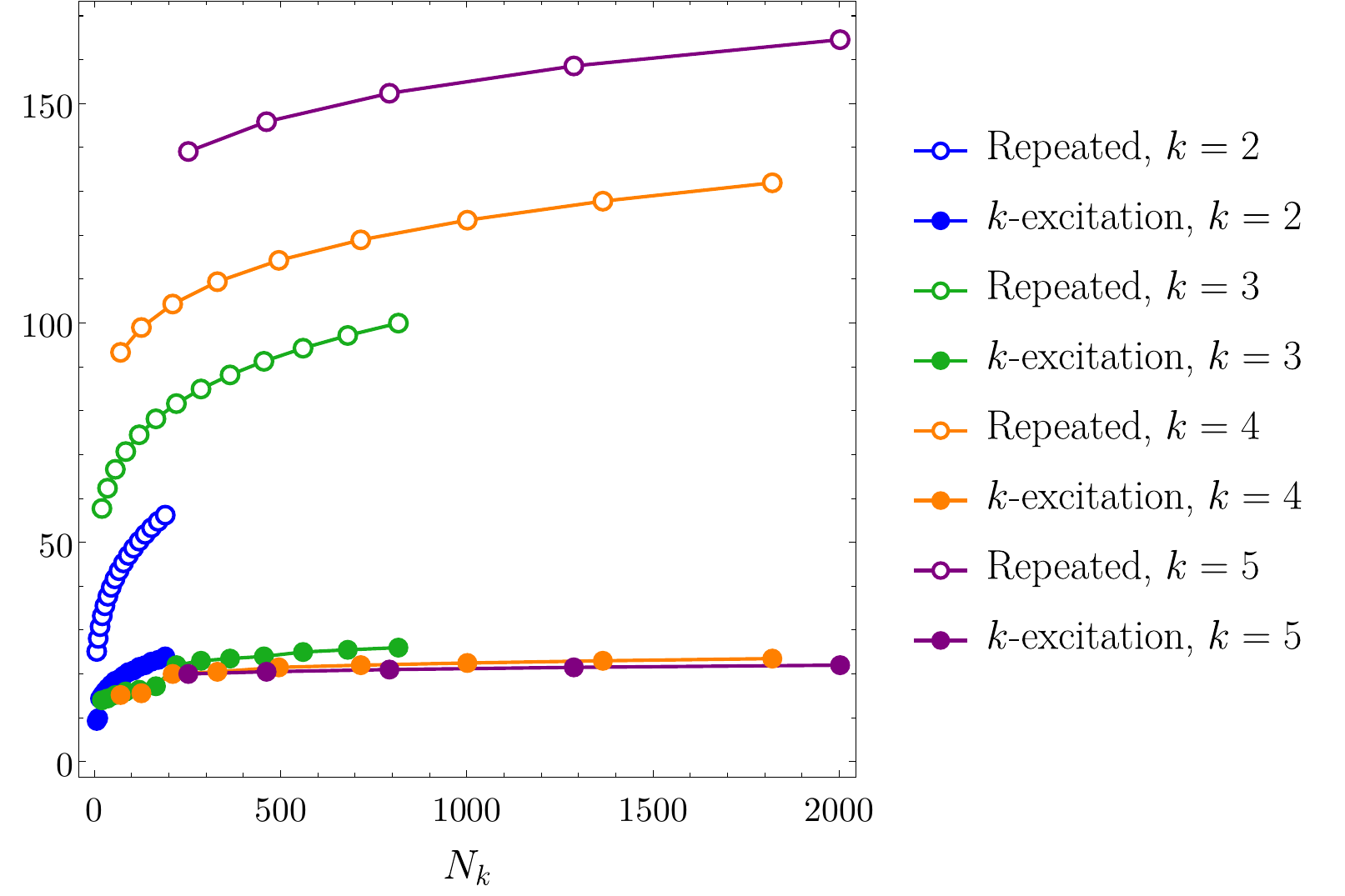}}
    \caption{Plot of the time to reach $P_\textrm{success} > 0.99$ for a binary string with $k$ excitations ($k$ number of 1s) using the repeated single-excitation subspace spatial search protocol, giving time $t^{(k)}_{\textrm{1-subspace}}$ of Eq.~\eqref{eq:time_1_subspace} (hollow circles), and the $k$-excitation spatial search protocol introduced here, giving time $t^{(k)}_{k\textrm{-subspace}}$ of Eq.~\eqref{eq:time_k_subspace} (solid circles). The same data is represented with respect to (a) the number of spins $n$, and (b) the size of the state space $N_k$ from $n$ spins. }
    \label{fig:time_to_reach_P_success}
\end{figure*}

We consider a comparison to a similar situation with $k$ marked sites but repeatedly performing the search in the single-excitation subspace. In this case, the final state after the continuous-time quantum walk is a superposition of states with a single 1 at each of the $k$ marked sites~\cite{chakraborty_spatial_2016}. We then randomly measure one of the $k$ marked sites. The idea is to find the number of successful trials of spatial search required to ensure we have found all $k$ marked sites and can reveal the target binary string. This is equivalent to the coupon collector's problem~\cite{ferrante_coupon_2014}, for which the expected number of trials is well known. Let $T$ be the number of trials of spatial search required to find all the marked sites and is a random variable. The expected number of trials to find the $k$ marked sites is
\begin{align}
    \mathbb{E}[T] &= k \sum_{i=1}^{k} \frac{1}{i} \\ &= k \log k + \gamma_\mathrm{EM} k + \frac{1}{2} + O(k^{-1}),
\end{align}
where $\gamma_{\mathrm{EM}} \approx 0.577$ is the Euler-Mascheroni constant. The number of trials to find the target binary string is therefore $O(k\log k)$. We want to establish the number of trials $s_k$ to have a greater than $1 - \varepsilon_s$ probability of finding all the $k$ marked sites. 
$P(T > \tau)$ is the probability of requiring at least $\tau$ trials to find all the marked sites. We need the smallest $s_k$ such that $P(T > s_k) < \varepsilon_s$, which is the probability that after $s_k$ trials at least one of the $k$ marked sites is missing, 
\begin{align}
    P(T > s_k) = \sum_{i=1}^{k}(-1)^{i+1}\binom{k}{i}\left(\frac{k-i}{k}\right)^s < \varepsilon_s.
\end{align}
As an example, for $\varepsilon_s = 0.01$, we numerically find $s_1=1$, $s_2=8$, $s_3=15$, $s_4=21$, and $s_5=28$. We must also consider the fidelity of the spatial search itself to find the total time required to reach a minimum fidelity. Given a maximum fidelity of spatial search $F$, we can repeat the spatial search $r_k$ times to find the marked binary string with higher fidelity. We can assume that the probability of the correct binary string is by far the most likely binary string. In this case, we do not require a majority vote, we only require the correct binary string to be seen twice. To achieve an overall fidelity of greater than $1-\varepsilon_r$, we require $r_k$ trials such that
\begin{align}
     \left(1 - F\right)^{r_k} + r_k
    F \left(1 - F\right)^{r_k-1} &< \varepsilon_r
\end{align}
Overall, the time for finding the $k$ marked sites with spatial search with error $\varepsilon = \varepsilon_s + \varepsilon_r - \varepsilon_s\varepsilon_r \approx \varepsilon_s + \varepsilon_r$ in the single-excitation subspace is 
\begin{equation}
    t^{(k)}_\textrm{1-subspace} = s_k r_k \frac{\pi \sqrt{n}}{2}.
    \label{eq:time_1_subspace}
\end{equation}
Similarly the time for spatial search in the $k$-excitation subspace is 
\begin{equation}
        t^{(k)}_{k\textrm{-subspace}} = r_k t^{(k)}_n,
        \label{eq:time_k_subspace}
\end{equation}
where $t^{(k)}_n$ is the time to reach the maximum fidelity for a particular $n$, so $t^{(k)}_n \sim \sqrt{n}$. 
We can now compute the time required to reach a given success probability $P_\textrm{success}$ of finding all $k$ marked sites. Our spatial search in higher-excitation subspaces protocol is compared to repeated spatial search in the single-excitation subspace. In both cases, we repeat the procedure such that $P_\textrm{success} > 0.99$, see Fig.~\ref{fig:time_to_reach_P_success}. Ignoring the fidelity decreasing in the higher-excitation subspace search, we would expect an $O(k\log k)$ separation in time due to $s_k$. We find that $ t^{(k)}_\textrm{1-subspace} \approx k(1+0.3 \log k) t^{(k)}_{k\textrm{-subspace}}$, which means a factor $k(1+0.3 \log k)$ advantage by using the $k$-excitation spatial search over the repeated spatial search.

\section{\label{sec:classical}Computational complexity}
The time to reach the maximum fidelity for $\gamma = 1/n$ is $O(\sqrt{n})$. The state space actually grows faster than linear for $k\geq 2$ -- since $N_k = O(n^k)$. This may seem to violate arguments for the fastest possible search of an unstructured database, which would suggest search in $O(\sqrt{N_k})$ time~\cite{grover_fast_1996,boyer_tight_1998}. However, the search we have described is structured. The oracle Hamiltonian ($H_\textrm{mark}$) does not only indicate the solution state (marked binary string) in the database, it also indicates the number of 1s in the query binary string that match the 1s in the marked binary string. Physically, the number of matching excitations affects the energy of the state through $H_\textrm{mark}$. The Hamiltonian applies a structure in the database by giving an effective distance -- related to the Hamming distance -- to the solution. In this way, we can still have an asymptotic time $t^{(k)}_\infty = O(\sqrt{n})$ to find the marked binary string for a state space of size $O(n^k)$. 

The quantum spatial search algorithm with multiple excitations in the $k$-excitation subspace is essentially a continuous-time analogue of implementing a Hamming distance oracle for structured search. The Hamming distance oracle returns the Hamming distance between the target binary string $w$ and the query binary string $a$, $d(w,a)$, giving the number of bits that are not in the correct locations. A Hamming distance oracle (for an oracle that returns the Hamming distance modulo 4) has been proven to solve the binary string search with a single query~\cite{hunziker_quantum_2002}. However, the complexity of the oracle itself must also be considered for a comparison to our case. Calculating the Hamming distance between two binary strings requires a linear bit-wise comparison, so it would give $O(\sqrt{n})$ time if amplitude amplification is possible, which would be equivalent to our result. 

The simplest classical algorithm would be a bit-wise search algorithm, checking whether each of the $n$ bits are correct, which would be $O(n)$ time. This does not use the information that there are $k$ 1s. The information contained in a binary string of length $n$ with $k$ 1s is $\log_2 N_k $ since there are $N_k = \binom{n}{k}$ possible configurations of binary strings with $k$ 1s. The information-theoretic lower bound for query complexity with an oracle that returns a single bit of information is therefore $\Omega(\log N_k)$. The Hamming distance oracle returns $\log_2(k)$ bits of information, which gives a information-theoretic lower bound $\Omega(\log N_k/\log{k})$, and computing the Hamming distance would require $O(n)$ time. Together, this would give an asymptotic classical complexity slower than that of bit-wise checking. 

The final relevant point concerning the computational complexity of the search algorithm is constructing the initial state. Generating the equal superposition of a subset of basis states, for example an equal superposition of a particular excitation subspace, is efficient in time $O(\log N_k) = O(k \log n)$  for the $k$-excitation subspace~\cite{shukla_efficient_2024}.

\section{\label{sec:conclusion}Conclusion}
We have presented and analysed the problem of finding a binary string with $k$ 1s that is marked by only local fields at each site that corresponds to a $1$. We use a continuous-time quantum walk, equivalent to the case of spatial search, but now in higher-excitation subspaces, giving a structured search. The Hamiltonian becomes an analogue implementation of a Hamming distance oracle that can find the solution with finite fidelity in time $O(\sqrt{n})$ for a state space size of $O(n^k)$. We use combinatorics to find an expression for the Hamiltonian as a $(k+1) \times (k+1)$ matrix and solve the asymptotic fidelity of the search for $k=2$ and $k=3$. The spin formalism of the problem is also introduced and solved, which confirm the results. Our low-dimensional Hamiltonian allows us to numerically compute the asymptotic fidelity for up to $k=150$. In order to demonstrate that this search protocol could be implemented on current ion trap platforms, we show that the algorithm works for long-range interactions in a 1-dimensional spin chain. The interaction range is parameterised by $\alpha$ and a range of up to $\alpha = 1.5$ still demonstrates $k$-excitation subspace spatial search with high fidelity for low $n$. 

The advantage of the protocol for finding binary strings with $k$ 1s is twofold for continuous-time quantum algorithms. Firstly, for a limited number of qubits, the state space is much larger in the $k$-excitation subspace rather than the single-excitation subspace. Thus, for a target number of states, it is more space efficient (in terms of number of qubits). In ion trap systems, there is a clear advantage in keeping the number of ions low, because lower $\alpha$ is possible, which gives higher fidelity spatial search. Secondly, the total run-time of the protocol to find the marked binary string is a factor of $O(k \log k)$ faster for $k$-excitation subspace spatial search over a repeated single-excitation spatial search. 

This algorithm effectively gives a quantum QR code reader, for low-density QR codes with predetermined $k$. The black (or white) parts of the quantum QR code can be represented by local fields. Or the QR code itself could be a lattice of spins. Each spin being in spin state up or down giving the black and white of the QR code. Bringing the ion trap lattice close to these spins produces local fields that could be detected quickly, in $O(\sqrt{n})$ time using this $k$-excitation spatial search protocol. In a similar way, the ion trap lattice could be a fast sensor for spin or magnetic defects.
Other continuous-time quantum algorithms could be performed in higher-excitation subspaces more generally or utilise the analogue implementation of the Hamming distance oracle that we have introduced. 

\begin{acknowledgements}
DL thanks Robert R Lewis for pointing out the relevant graphs are called Johnson graphs. DL acknowledges support from the EPSRC Centre for Doctoral Training in Delivering Quantum Technologies, grant ref. EP/S021582/1.
LB acknowledges financial support from: PNRR Ministero Università e Ricerca Project 
No. PE0000023-NQSTI funded by European Union-Next-Generation EU. 
Prin 2022 - DD N.~104 del 2/2/2022, entitled ``understanding the LEarning process of QUantum
Neural networks (LeQun)'', proposal code 2022WHZ5XH, CUP B53D23009530006;
U.S. Department of Energy, Office of Science, National Quantum Information Science Research
Centers, Superconducting Quantum Materials and Systems Center (SQMS) under the Contract
No. DE-AC02-07CH11359. 
\end{acknowledgements}

\onecolumngrid
\appendix
\section*{Appendix}
\section{\label{app:spatial_multiple_excitations}Spatial search in higher-excitation subspaces}
\begin{figure*}[htb]
    \centering
    \subfloat[$J(4,2)$]{
    \includegraphics[scale=0.74]{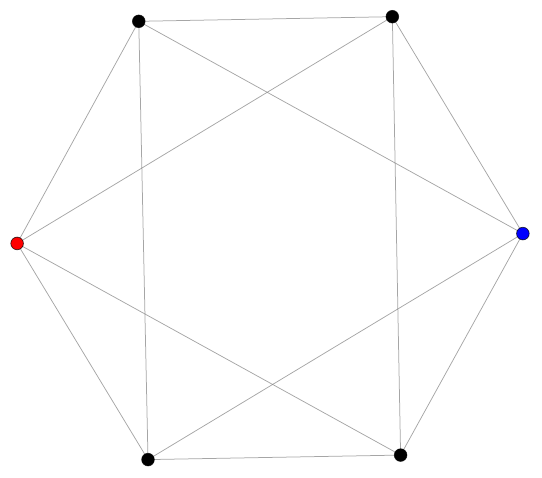}}
    \subfloat[$J(6,2)$]{\includegraphics[scale=0.74]{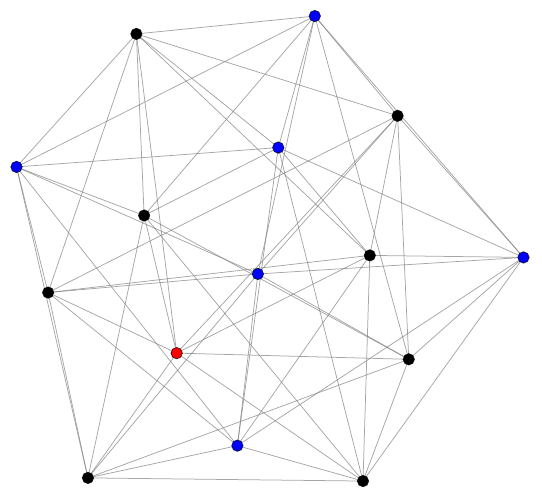}}
    
    \subfloat[$J(8,2)$]{\includegraphics[scale=0.74]{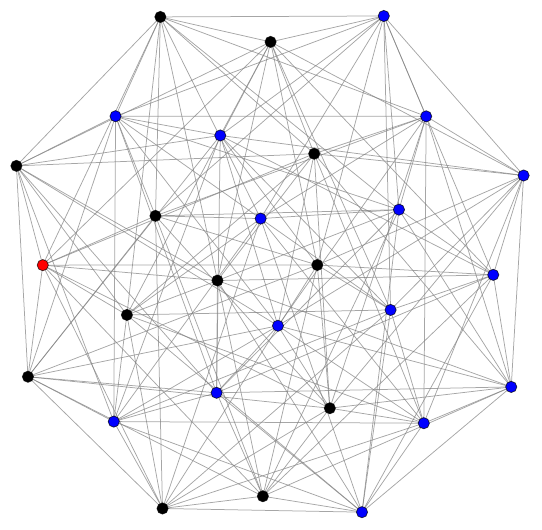}}
    \subfloat[$J(8,3)$]{\includegraphics[scale=0.74]{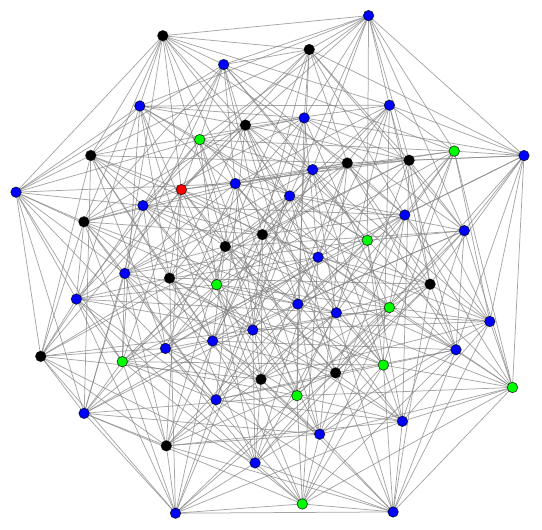}}
    \caption{Various Johnson graphs are depicted. There are two or three excitations in the states of these graphs (there are two or three 1s and the rest of the spin states are 0). The red vertex indicates the state $|w\rangle$; the black vertices are distance 1 from $w$; the blue vertices are distance 2 from $w$; and the green vertices are distance 3 from $w$. Equivalently, the red vertex is the state where all excitations of the spin state are precisely on local fields; the black vertices are the sites where a single excitation of the spin state is not on a local field; the blue vertices are where two excitations of the spin state are not on the local fields; and the green vertices are where three excitations of the spin state are not on the local fields.}
    \label{fig:johnson_graphs_Jn2}
\end{figure*}

In this section, we introduce the relevant notation and construct the search Hamiltonian in the reduced subspace of size $(k+1) \times (k+1)$. 
As described in Sec.~\ref{sec:preliminaries}, the Johnson graph $J(n,k)$ is the walk Hamiltonian of $k$-excitation subspace search with $n$ spins. In Fig.~\ref{fig:johnson_graphs_Jn2}, we show various Johnson graphs for the two- and three-excitation spin subspaces, $k=2$ and $k=3$.

The graph $J(n,k)$ with $k\leq n$ has degree $k(n-k)$ and diameter $k$. For $k=2$, the graph has $N_2 = \frac{1}{2}n(n-1)$ vertices, degree $2(n-2)$ and diameter $2$. This means that each vertex is at most a distance $2$ from the vertex corresponding to the marked state. We define the sets of vertices $\mathcal{D}_{k,q}^{(a)} = \set{v\in\mathcal{V}_k | d(a,v) = 2q }$, which is the set of vertices a distance $q$ from a reference vertex $a$ in the graph. We define $d_{k,q} = \vert \mathcal{D}_{k,q}^{(w)}\vert$ as the number of vertices that are a graph distance $q$ away from the vertex of state $|w\rangle$. The number of vertices a distance 1 away from the marked state is the degree. For $k=2$, we have $d_{2,1} = 2(n-2)$ and the number of vertices a distance 2 away is therefore $d_{2,2} = N_2 -d_2 - 1 = \frac{1}{2}(n-2)(n-3)$. For general $k$ and $q$, we can compute $d_{k,q}$ using straightforward combinatorics of binary strings. The number of 1s originally is $k$. These 1s must become $k-q$ 1s and $q$ 0s to be in the set $\mathcal{D}^{(w)}_{k,q}$. This gives $\binom{k}{q}$ combinations. The number of 0s originally is $n-k$. These 0s must become $q$ 1s and $n-k-q$ 0s. This is $\binom{n-k}{q}$ combinations. Overall, we have 
\begin{align}
    d_{k,q} = \binom{k}{q}\binom{n-k}{q}.
\end{align}

In general, we can consider the state $|d_{k,q}\rangle$, which is an equal superposition of states corresponding to the vertices a distance $q$ away from the vertex of state $|w\rangle$,
\begin{align}
    |d_{k,q}\rangle = \frac{1}{\sqrt{d_{k,q}}} \sum_{a \in \mathcal{D}_{k,q}^{(w)}} |a\rangle. 
\end{align}
A subspace of the graph can be considered with reference to the marked spin: the marked state $|w\rangle$ itself, a superposition state of vertices a distance 1 away from the marked state $|d_{k,1}\rangle$, and a superposition state of the vertices a distance 2 away from the marked state $|d_{k,2}\rangle$, etc.. For $k$ excitations, we require state $|d_{k,k}\rangle$ since the diameter of the graph is $k$. This means the subspace size is $k+1$. The colours of the vertices, red, black and blue, in Fig.~\ref{fig:johnson_graphs_Jn2} correspond to vertices in the sets $\{w\}$, $\mathcal{D}^{(w)}_{2,1}$, $\mathcal{D}^{(w)}_{2,2}$ respectively. 
 The general superposition state can be written 
\begin{align}
    |s_k\rangle = \frac{1}{\sqrt{N_k}} \sum_{a \in \mathcal{V}_k} |a\rangle,
\end{align}
which can therefore be expressed with the subspace superposition states as 
\begin{align}
    |s_k\rangle = \frac{1}{\sqrt{N_k}} \left(|w\rangle + \sum_{q=1}^{k} \sqrt{d_{k,q}}|d_{k,q}\rangle\right).
\end{align}

Since the marked Hamiltonian is diagonal it can be written in the subspace as Eq.~\eqref{eq:mark_Hamiltonian_strings}. As an example, for the 2-excitation subspace, we have the marked Hamiltonian, 
\begin{align}
    H_\textrm{mark} = |w\rangle\langle w| - |d_{2,2}\rangle\langle d_{2,2}|.
\end{align}
This can be compared to the spatial search Hamiltonian of the single-excitation subspace, which is $H_\textrm{mark} = |w\rangle\langle w|$. The walk Hamiltonian is more complex. The initial state is the superposition state $|s_k\rangle$ and is an eigenstate of $H$ -- the ground state. Thus, since $H_\textrm{mark}$ is diagonal in the subspace $\{|w\rangle, |d_{2,1}\rangle, |d_{2,2}\rangle\}$, the evolution of the state $|s_k\rangle$ can be completely characterised in this subspace. 

\begin{figure}
    \centering
    \includegraphics[scale=0.5]{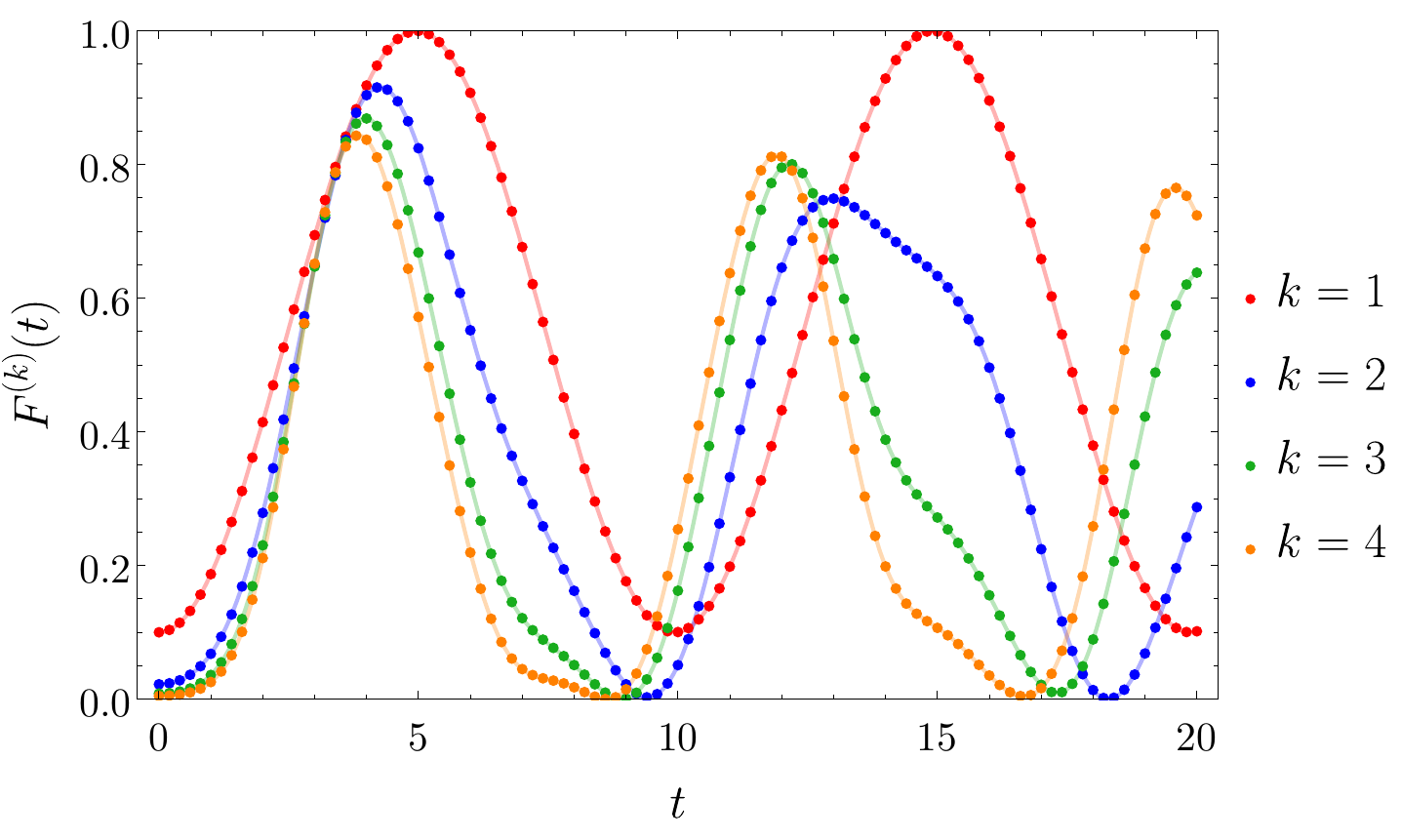}
    \caption{Plot of $F^{(k)}(t)$ for $\gamma=0.1$, $n=10$ and various $k$. The points are from the simulation of the full Hamiltonian evolution. The solid line is the simulation of the evolution using the subspace dynamics with the Hamiltonian of Eq.~\eqref{eq:subspace_hamiltonian}.}
    \label{fig:compare_full_and_subspace_evolution}
\end{figure}

Two sets are important for the walk Hamiltonian matrix elements: firstly, the set of distance one vertices from a vertex $a$, $\mathcal{D}^{(a)}_{k,1}$, where $a \in \mathcal{D}^{(w)}_{k,i-1}$ is a vertex a distance $i-1$ from the marked vertex $w$; and secondly, the set of vertices a distance $j-1$ from $w$, which is $\mathcal{D}^{(w)}_{k,j-1}$. The size of the intersection of these two sets is multiplied by the ratio of their state normalisations to give the Hamiltonian matrix element $H_{ij}$. The reason the normalisations of the state are required is that the Hamiltonian cannot increase (or decrease) the total amplitude that is transferred from one subspace to another subspace. We use combinatorics to calculate the size of the set intersections,
\begin{align}
    \label{eq:intersection_sets}
    \left\vert  \mathcal{D}_{k,1}^{(a)} \cap \mathcal{D}_{k,j-1}^{(w)} \right\vert \textrm{ for } a \in \mathcal{D}_{k,i-1}^{(w)} = \begin{cases}
         (i-1)(n-2i+2) & \textrm{for } i=j,\\
         (k-i+1)(n-k-i+1) & \textrm{for } j-i=1,\\
         0 & \textrm{otherwise},
    \end{cases} 
\end{align}
where we have assumed $j\geq i$ since we know the Hamiltonian is symmetric. We also define $\mathcal{D}_{k,0}^{(w)} = \{w\}$ and $d_{k,0} = 1$. The combinatorics that produces this result is given in Appendix~\ref{app:combinatorics}.  
Thus, we have 
\begin{align}
    H_{i j} &=  \frac{\sqrt{d_{k,i-1}}}{\sqrt{d_{k,j-1}}} \times \begin{cases}
         (j-1)(n-2j+2) & \textrm{for } i=j,\\
         (k-i+1)(n-k-i+1) & \textrm{for } j-i=1,\\
         0 & \textrm{otherwise},
    \end{cases} 
    \\
    &= \begin{cases}
         (j-1)(n-2j+2) & \textrm{for } i=j,\\
         i\sqrt{(k-i+1)(n-k-i+1)} & \textrm{for } j-i=1,\\
         0 & \textrm{otherwise}.
    \end{cases} 
\end{align}
The result for the initial state $|s_k\rangle$ is exact. We define the fidelity, 
\begin{align}
    F^{(k)}(t) = \vert \langle w | e^{-i H_\textrm{search}t }| s_k \rangle \vert^2.
\end{align}
In Fig.~\ref{fig:compare_full_and_subspace_evolution}, we compare the fidelity for the subspace evolution with that of the full Hamiltonian dynamics with the marking and walk Hamiltonians from Eqs.~\eqref{eq:mark_Hamiltonian} and~\eqref{eq:walk_Hamiltonian_spin} respectively.

\subsection{\label{app:two_excitation_subspace}2-excitation subspace}
The 2-excitation subspace is given by $k=2$. The walk Hamiltonian, with a hopping rate $\gamma$, and a marked Hamiltonian gives the total search Hamiltonian
\begin{align}
    H_\textrm{search} = \begin{pmatrix}
        1 & \gamma\sqrt{2(n-2)} & 0 \\
        \gamma\sqrt{2(n-2)} & \gamma(n-2) & 2\gamma\sqrt{n-3} \\
        0 & 2\gamma\sqrt{n-3} & 2\gamma(n-4) - 1
    \end{pmatrix}.
		\label{eq:H search 2}
\end{align}
We can set $\gamma = 1/(n-2)$, to give 
\begin{align}
    H_\textrm{search} = \begin{pmatrix}
        1 & \frac{\sqrt{2}}{\sqrt{n-2}} & 0 \\
        \frac{\sqrt{2}}{\sqrt{n-2}} & 1 & 2\frac{\sqrt{n-3}}{n-2} \\
        0 & 2\frac{\sqrt{n-3}}{n-2} & 1-\frac{4}{n-2}
    \end{pmatrix}.
\end{align}
For large $n$, we have the first order expansion
\begin{align}
     H_\textrm{search} &= \begin{pmatrix}
        1 & \frac{\sqrt{2}}{\sqrt{n}} & 0 \\
        \frac{\sqrt{2}}{\sqrt{n}} & 1 & \frac{2}{\sqrt{n}} \\
        0 & \frac{2}{\sqrt{n}} & 1
    \end{pmatrix} + O(n^{-1}),\\
    &= \mathds{1} + \frac{1}{\sqrt{n}}R_2 + O(n^{-1}),
\end{align}
where we have defined 
\begin{align}
    R_2 = \begin{pmatrix}
        0 & \sqrt{2} & 0 \\
        \sqrt{2} & 0 & 2 \\
        0 & 2 & 0
    \end{pmatrix}.
\end{align}
which gives a first order evolution of 
\begin{align}
    U_0(t) &= e^{-i t} e^{-i R_2 \frac{t}{\sqrt{n}}} \\
    &= \frac{e^{-i t}}{3} \begin{pmatrix}
        2+ \cos(\tfrac{\sqrt{6}t}{\sqrt{n}
        }) & i \sqrt{3} \sin(\tfrac{\sqrt{6}t}{\sqrt{n}
        }) & \sqrt{2}\left( \cos(\frac{\sqrt{6}t}{\sqrt{n}}) -1 \right) \\
        i \sqrt{3} \sin(\tfrac{\sqrt{6}t}{\sqrt{n}
        }) & 3 \cos(\tfrac{\sqrt{6}t}{\sqrt{n}
        }) & i \sqrt{6} \sin(\tfrac{\sqrt{6}t}{\sqrt{n}
        }) \\
        \sqrt{2}\left( \cos(\frac{\sqrt{6}t}{\sqrt{n}}) -1 \right) &  i \sqrt{6} \sin(\tfrac{\sqrt{6}t}{\sqrt{n}
        }) & 1+ 2\cos(\tfrac{\sqrt{6}t}{\sqrt{n}
        }).
    \end{pmatrix}
\end{align}
The asymptotic fidelity $F_\infty^{(2)}$ is defined as the first order expansion about large $n$ of the fidelity $F^{(2)} = \vert \langle w | e^{-iH_\textrm{search} t} | s \rangle \vert^2$ for finding the marked state for $k=2$. This is 
\begin{align}
    F_\infty^{(2)}(t) = \frac{4}{9}\left(\cos(\tfrac{\sqrt{6} t}{\sqrt{n}}) -1\right)^2 ,
\end{align}
where we have used the asymptotic expressions $d_{2,1} \sim 2n$, $d_{2,2} \sim n^2/2$, and $N_2 \sim n^2/2$. The maximum asymptotic fidelity is therefore $|F_\infty^{(2)}| = \max_{t\in \mathbb{R}^+}\{F_\infty^{(2)}(t)\} = \frac{8}{9}$ and occurs at a time $t^{(2)}_\infty/\sqrt{n} = \frac{\pi\sqrt{n} }{\sqrt{6}}$. Crucially, this time scales with the number of spins as $\sqrt{n}$ despite a state space size that grows quadratically, $N_2 = \frac{1}{2}n(n-1)$. 
\subsection{\label{app:three_excitation_subspace}3-excitation subspace}
The 3-excitation subspace is given by $k=3$. The marked Hamiltonian is 
\begin{align}
    H_\textrm{mark} = \frac{3}{2}|w\rangle\langle w| + \frac{1}{2}|d_{3,1}\rangle\langle d_{3,1}| - \frac{1}{2}|d_{3,2}\rangle\langle d_{3,2}| - \frac{3}{2} |d_{3,3}\rangle\langle d_{3,3}|. 
\end{align}
The walk Hamiltonian, with a hopping rate $\gamma$, and the marked Hamiltonian above gives the total search Hamiltonian
\begin{align}
    H_\textrm{search} = \begin{pmatrix}
        \frac{3}{2} & \gamma\sqrt{3(n-3)} & 0 & 0 \\
        \gamma\sqrt{3(n-3)} &  \gamma(n-2) + \frac{1}{2} & 2\gamma\sqrt{2(n-4)} & 0 \\
        0 & 2\gamma\sqrt{2(n-4)} & 2\gamma(n-4) - \frac{1}{2} & 3\gamma \sqrt{n-5} \\
        0 & 0  & 3\gamma \sqrt{n-5}  & 3\gamma (n-6)  - \frac{3}{2}.
    \end{pmatrix}.
		\label{eq:H search 3}
\end{align}
We consider the case of $\gamma = 1/n$ for large $n$. All the diagonal components of $H_\textrm{search}$ tend to $3/2$ as $n$ becomes large. As in the $k=2$ case, we consider the first order expansion about large $n$, 
\begin{align}
    H_\textrm{search} = \frac{3}{2}\mathds{1} + \frac{1}{\sqrt{n}}R_3 + O(n^{-1}),
\end{align}
with
\begin{align}
    R_3 = \begin{pmatrix}
        0 & \sqrt{3} & 0 & 0\\
        \sqrt{3} & 0 & 2\sqrt{2} & 0 \\
        0 & 2\sqrt{2} & 0 & 3 \\
        0 & 0 & 3 & 0
    \end{pmatrix}.
\end{align}
We only have to consider the time evolution of the off-diagonal elements $R_3$. In the asymptotic limit of large $n$, the initial state becomes $\lim_{n\rightarrow\infty}|s_3\rangle = |d_{3,3}\rangle$ and the final state remains $|w\rangle$. In the subspace basis, these states can be labelled $|4\rangle$ and $|1\rangle$ respectively. The asymptotic fidelity is therefore 
\begin{align}
    F_\infty^{(3)}(t)&= \big\vert\langle 1| e^{-i \frac{R_3}{\sqrt{n}} t} |4\rangle \big\vert^2 \\
    &= \frac{2}{73} \left( \sqrt{10 + \sqrt{73}} \sin(\tfrac{\sqrt{10-\sqrt{73}}}{\sqrt{n}} t) - \sqrt{10 - \sqrt{73}}\sin(\tfrac{\sqrt{10+\sqrt{73}}}{\sqrt{n}}t) \right)^2 .
\end{align}
Finding $F_\infty^{(3)}(t)$ at $t=\frac{\pi \sqrt{n} }{2 \sqrt{10-\sqrt{73}}}$, we find $\vert F^{(3)}_\infty \vert \geq 0.702$. Again, as in the $k=2$ subspace, we find a finite asymptotic fidelity in a time that scales with the number of spins as $t_\infty^{(3)} = O(\sqrt{n})$. The state space now grows as $N_3 = \frac{1}{6} n(n-1)(n-2)= O(n^3)$ with the number of spins.

\section{\label{app:combinatorics}Combinatorics}
In this section, we explain in detail how the result of Eq.~\eqref{eq:subspace_hamiltonian} can be found. As explained in App.~\ref{app:spatial_multiple_excitations}, the Hamiltonian elements are 
\begin{align}
    H_{ij} = \frac{\sqrt{d_{k,i-1}}}{\sqrt{d_{k,j-1}}} \times \left\vert  \mathcal{D}_{k,1}^{(a)} \cap \mathcal{D}_{k,j-1}^{(w)} \right\vert \textrm{ for } a \in \mathcal{D}_{k,i-1}^{(w)}.
\end{align}
We now describe the combinatorics to compute the Hamiltonian elements. In particular, we show how the following result is obtained,
\begin{align}
    \left\vert  \mathcal{D}_{k,1}^{(a)} \cap \mathcal{D}_{k,j-1}^{(w)} \right\vert \textrm{ for } a \in \mathcal{D}_{k,i-1}^{(w)} = \begin{cases}
         (i-1)(n-2i+2) & \textrm{for } i=j,\\
         (k-i+1)(n-k-i+1) & \textrm{for } j-i=1,\\
         0 & \textrm{otherwise}.
    \end{cases} 
\end{align}
For convenience, we restate the definition of the sets $\mathcal{V}_k = \set{ b \in \mathcal{V} | \sum_i^n b_i = k} $, with $b_i$ the value of the $i$th bit of $b$, and $\mathcal{D}^{(a)}_{k,q} = \set{v\in\mathcal{V}_k | d(a,v) = 2q }$, with $d(a,b)$ being the Hamming distance between binary strings $a$ and $b$. $\mathcal{V}_k$ is the set of binary strings with $k$ excitations (1s). $\mathcal{D}^{(a)}_{k,q}$ is the set of vertices a graph distance $q$ from a vertex $a$ in the graph of binary strings with $k$ excitations (1s). 

Firstly, we note that the graph is vertex transitive with all-to-all interactions, which means the specific choice of marked vertex makes no difference.
To explain the combinatorics, we take the case that the marked vertex is $w = 1\dots 10\dots 0$, which we can represent as $w = \newmoon \dots \newmoon \fullmoon \dots \fullmoon$.
Every spin is indistinguishable with all-to-all interactions between spins, which means the specific choice of marked vertex makes no difference. 

We consider three cases separately. First, the case $i=j$. In this case, we consider the set of vertices $\mathcal{D}_{k,i-1}^{(w)}$ that are a graph distance $i-1$ from $w$. To construct vertices a distance $i-1$ from $w$, we can start with the $w$ string and swap $i-1$ of the total number of $\newmoon$ with $i-1$ of the $n-k$ total number of $\fullmoon$, an example of the swaps are
\begin{align}
    \newmoon \dots \newmoon \fullmoon \dots \fullmoon \rightarrow   \underbrace{\newmoon \dots \newmoon}_{A} \overbrace{\fullmoon \dots \fullmoon}^{B} \underbrace{\newmoon \dots \newmoon}_{C} \overbrace{\fullmoon \dots \fullmoon}^{D},
    \label{eq:ABCD_combinations}
\end{align}
where $A$ is the $k-(i-1)$ number of $\newmoon$ that are not swapped, $B$ is due to the $i-1$ number of $\fullmoon$ that are swapped, $C$ is the $i-1$ number of $\newmoon$ due to the swap, and $D$ is the $n-k-(i-1)$ number of $\fullmoon$ not involved in the swap. 
The question is then how many binary strings are in this set that are also a graph distance 1 from this set. The answer is that we can (a) take any $\fullmoon$ in $B$ and swap it with a $\newmoon$ in $A$, or (b) we can take any $\newmoon$ in $C$ and swap it with a $\fullmoon$ in $D$. 
The number of swaps possible are the number of swaps for (a), which is $(i-1)(k-(i-1))$, plus the number of swaps for (b), which is $(i-1)(n-k-(i-1))$, giving a total 
\begin{align}
    \left\vert  \mathcal{D}_{k,1}^{(a)} \cap \mathcal{D}_{k,i-1}^{(w)} \right\vert \textrm{ for } a \in \mathcal{D}_{k,i-1}^{(w)} &= (i-1)(k-(i-1)) + (i-1)(n-k-(i-1))\\
    &= (i-1)(n-2i+2).
\end{align}
The second case is when $j=i+1$. Here, as in Eq.~\eqref{eq:ABCD_combinations}, we have two situations: (i) where $A$ is size $k-(i-1)$ and $C$ is size $i-1$; and, (ii) where $A$ is size $k-i$ and $C$ is size $i$. The set we are considering is the binary strings a graph distance 1 from (i) intersecting with (ii). The only way for the binary strings to intersect is if the binary string a distance 1 from (i) gives a binary string in set (ii), i.e. the swap takes the graph from a distance $i-1$ to $i$ from $w$. Taking set (i), we therefore swap a $\fullmoon$ in $D$ with a $\newmoon$ in $A$ to increase the distance to $w$ by 1 and to now be in set (ii). The number of ways to do this is the number of $\newmoon$ in $A$, which is $k-(i-1)$, times the number of $\fullmoon$ in $D$, which is $n-k-(i-1)$, giving
\begin{align}
    \left\vert  \mathcal{D}_{k,1}^{(a)} \cap \mathcal{D}_{k,i}^{(w)} \right\vert \textrm{ for } a \in \mathcal{D}_{k,i-1}^{(w)} = (k-i+1)(n-k-i+1).
\end{align}
Finally, if $|i-j| > 1$, then the intersection is of sets of binary strings that are a different graph distance from $w$. Thus, it is not possible for there to be any intersection. Altogether, we find the result of Eq.~\eqref{eq:intersection_sets}, which gives Eq.~\eqref{eq:subspace_hamiltonian} in the main text.

\section{\label{app:spin_formalism}Spin formalism}

We now rewrite the spatial search problem in the language of spin physics. We introduce the following
three global spin operators
\begin{align}
	\vec S_1 &= \frac12 \sum_{i\in\mathcal M} \vec\sigma, &
	\vec S_2 &= \frac12 \sum_{i\notin\mathcal M} \vec\sigma, &
	\vec S &= \vec S_1+ \vec S_2. 
\end{align}
Then the walk and marking Hamiltonians, Eqs.\eqref{eq:walk_Hamiltonian_spin} and \eqref{eq:mark_Hamiltonian_strings}, 
for the fully connected graph $J_{ij}=1$ can be expressed as 
\begin{align}
	H&=  (S^x )^2 + (S^y )^2 - \frac n2 \openone = 
	\vec S\cdot\vec S - (S^z )^2 - \frac n2 \openone, \\
	 &=  S_1^+S_2^-+S_2^+S_1^-+\frac12\left(
	S_1^+S_1^-+S_2^+S_2^-+S_1^-S_1^++S_2^-S_2^+
	 \right) -\frac n2 \openone,
	\\
	H_{\rm mark} &= S_1^z,
\end{align}
where $S_j^\pm = S^x_j\pm i S^y_j$. 
Following the notation of standard quantum mechanics textbooks \cite{sakurai2020modern}, 
the walk Hamiltonian is diagonal in the basis $\ket{S,M}$ of the total spin operator $\vec S$, where $M$ is the eigenvalue 
of $S_z$, while the marking Hamiltonian is diagonal in $\ket{S_1,M_1,S_2,M_2}$, namely in the basis the operators 
$\vec S_1$ and $\vec S_2$, where $M_i$ is the eigenvalue of $M_i$. Since in $\mathcal V_k$ the number of excitations is fixed, 
so it is the total magnetization 
\begin{equation}
	M = \frac k2-\frac{n-k}2.
	\label{eq:total mag}
\end{equation}
while the magnetization of the marked and unmarked spins are related as $M_1+M_2=M$. 
The eigenvalues of search Hamiltonian are then 
$S(S+1)-M^2-n/2$. 

Since the entire Grover search protocol is symmetric upon exchanging spins
within both the marked and unmarked spaces, we can assume that $S_1$ and $S_2$ take their largest value 
\begin{align}
	S_1&=\frac k 2, & S_2 = \frac{n-k}{2},
	\label{eq:local spins}
\end{align}
while $M$ is fixed as \eqref{eq:total mag}. Indeed, the states with largest and smallest magnetization belong to the symmetric 
subspace, and since $-S_i\leq M_i\leq S_i$ this implies that $S_i$ has to be maximum. 
Restricting marked and unmarked spins to their symmetric subspaces, 
for $k\leq n/2$, the search Hamiltonian  can be expressed 
as a $(k{+}1)\times (k{+}1)$ matrix, since $M_1$ can take $k+1$ possible values. 
When $k>n/2$ we note that $M_2$ can take $n-k+1$ possible values, so we can extract 
an effective $(n-k+1)\times(n-k+1)$ Hamiltonian matrix. From the physical point of view, 
we can globally flip all spins without changing the spatial search
problem, resulting in the same $H$ but flipped $H_{\rm mark}$. 
Since we can always reabsorb the above operations by 
flipping the sign of $\gamma$, in what follows we focus on $k\leq n/2$. 

The spin operators act on the basis states $\ket{S,M}$ as
\begin{align}
	S^\pm \ket{S,M} &= \sqrt{S(S+1)-M(M\pm1)}\ket{S,M\pm1}, &
	S^z \ket{S,M} &= M\ket{S,M}. 
\end{align}
Setting $\ket{j_{n,k}} = \ket{S_1,M_1;S_2,M-M_1}$ where $M, S_1,S_2$ 
are given by Eqs.~\eqref{eq:total mag}-\eqref{eq:local spins}
and $M_1=S_1+1-j$, we get 
\begin{equation}
	H_{\rm search}^{n,k} = \sum_{j=1}^{k+1} \left(\gamma D_j +\frac k 2 +1 -j\right)
	\ket{j_{n,k}}\!\bra{j_{n,k}} + \gamma \sum_{j=1}^k \left(J_j 	\ket{j+1_{n,k}}\!\bra{j_{n,k}} + {\rm h.c.}\right),
	\label{eq:Hnk}
\end{equation}
where
\begin{align}
	J_j &= j \sqrt{k+1-j} \sqrt{n+1-k-j},
			&
	D_j &= (j-1) (n+2-2j).
\end{align}
The Hamiltonian $H_{\rm search}^{n,k}$ is equal to \eqref{eq:H search 2} for $k=2$ and to \eqref{eq:H search 3} for $k=3$. 

\section{\label{app:long_range_interaction_graphs}Fidelity with long-range interactions}
\begin{figure}[hb]
    \centering
    \subfloat[$n=10$]{\includegraphics[scale=0.4]{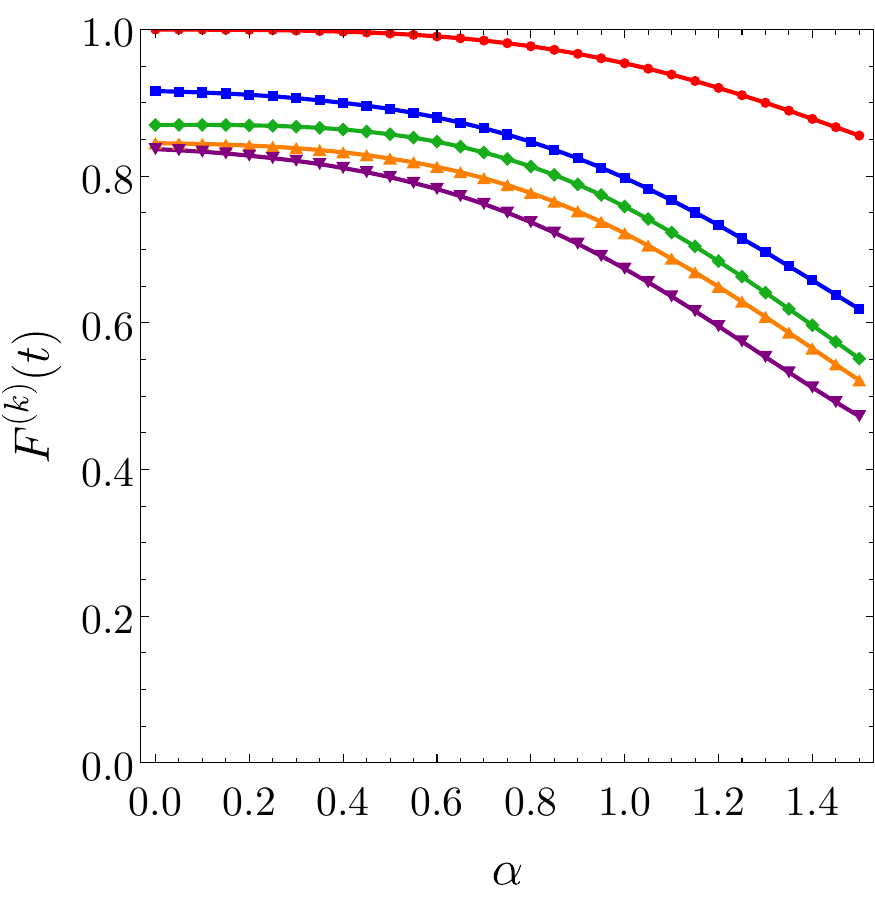}}
    \hspace{2mm}
    \subfloat[$n=11$]{\includegraphics[scale=0.4]{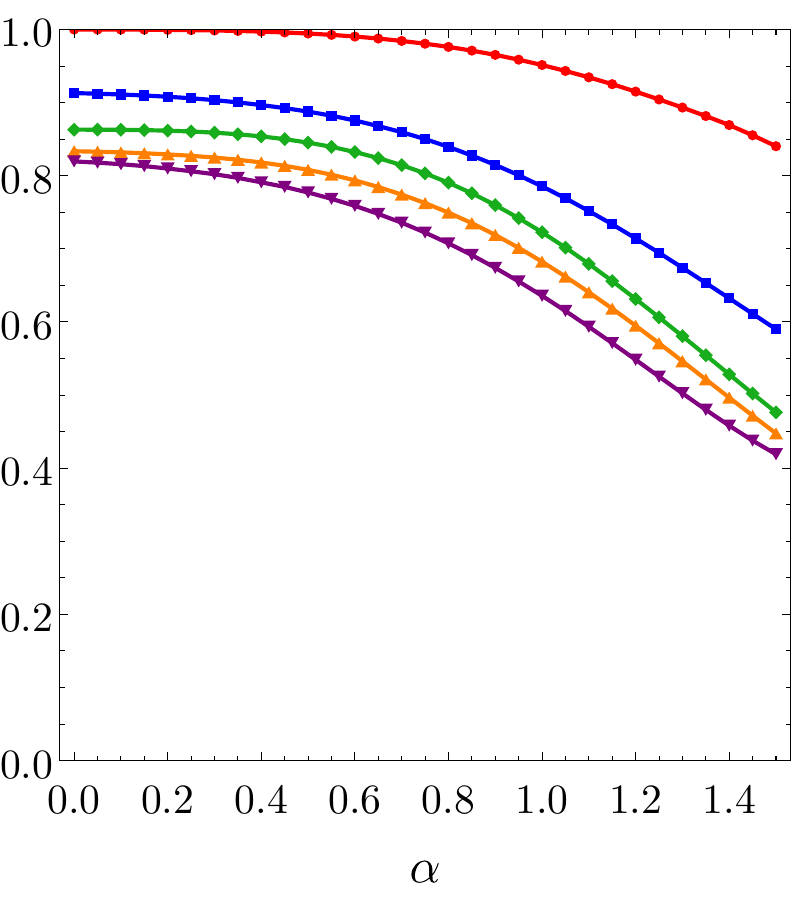}}
    \hspace{2mm}
    \subfloat[$n=12$]{\includegraphics[scale=0.4]{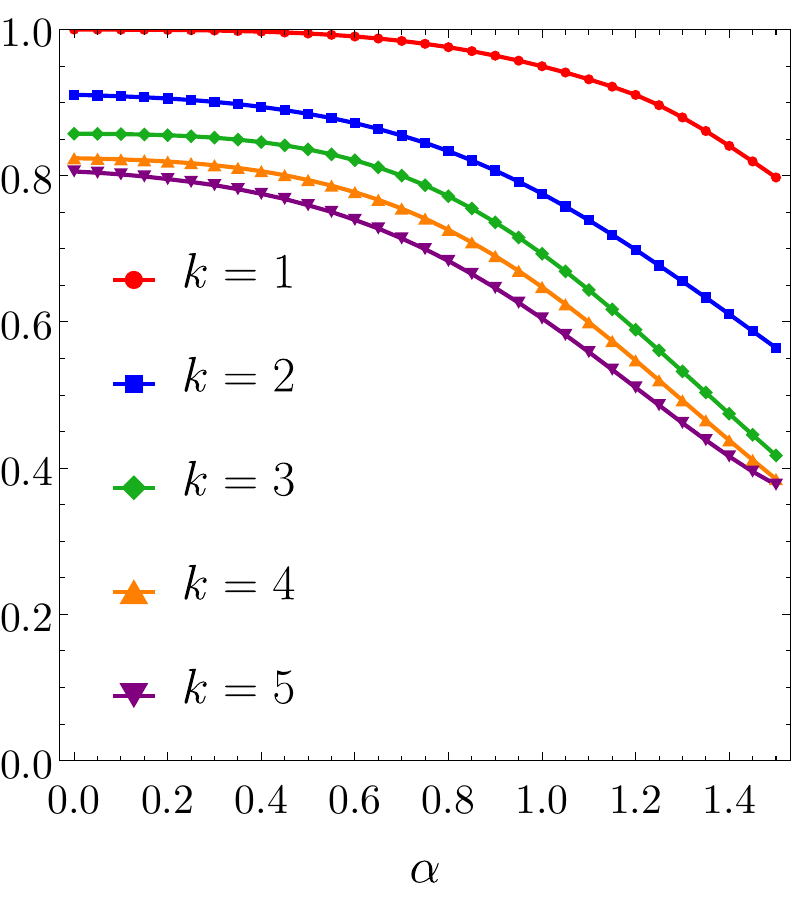}}
    \caption{Plot of $F^{(k)}(t) = |\langle w|e^{-i H_\textrm{search} t} |s\rangle|^2$ for varying interaction range, where $H_\textrm{search}$ coupling is defined in Eq.~\eqref{eq:long_range_coupling}, parameterised by $\alpha$. The numbers of spins $n$ are (a) n = 10, (a) n = 11, and (c) n = 12 with $k=1$ (red), $k=2$ (blue), $k=3$ (green), $k=4$ (orange), and $k=5$ (purple). The hopping rate $\gamma$ is optimised for each $\alpha$ using a line search.}
    \label{fig:fidelities_for_low_n_against_alpha}
\end{figure}

To complement Fig.~\ref{fig:fidelities_for_n_12_against_alpha} in the main text, we present the fidelity with long-range interactions for $n=10, 11, 12$ in Fig.~\ref{fig:fidelities_for_low_n_against_alpha}.


\begin{thebibliography}{26}%
	\makeatletter
	\providecommand \@ifxundefined [1]{%
		\@ifx{#1\undefined}
	}%
	\providecommand \@ifnum [1]{%
		\ifnum #1\expandafter \@firstoftwo
		\else \expandafter \@secondoftwo
		\fi
	}%
	\providecommand \@ifx [1]{%
		\ifx #1\expandafter \@firstoftwo
		\else \expandafter \@secondoftwo
		\fi
	}%
	\providecommand \natexlab [1]{#1}%
	\providecommand \enquote  [1]{``#1''}%
	\providecommand \bibnamefont  [1]{#1}%
	\providecommand \bibfnamefont [1]{#1}%
	\providecommand \citenamefont [1]{#1}%
	\providecommand \href@noop [0]{\@secondoftwo}%
	\providecommand \href [0]{\begingroup \@sanitize@url \@href}%
	\providecommand \@href[1]{\@@startlink{#1}\@@href}%
	\providecommand \@@href[1]{\endgroup#1\@@endlink}%
	\providecommand \@sanitize@url [0]{\catcode `\\12\catcode `\$12\catcode
		`\&12\catcode `\#12\catcode `\^12\catcode `\_12\catcode `\%12\relax}%
	\providecommand \@@startlink[1]{}%
	\providecommand \@@endlink[0]{}%
	\providecommand \url  [0]{\begingroup\@sanitize@url \@url }%
	\providecommand \@url [1]{\endgroup\@href {#1}{\urlprefix }}%
	\providecommand \urlprefix  [0]{URL }%
	\providecommand \Eprint [0]{\href }%
	\providecommand \doibase [0]{https://doi.org/}%
	\providecommand \selectlanguage [0]{\@gobble}%
	\providecommand \bibinfo  [0]{\@secondoftwo}%
	\providecommand \bibfield  [0]{\@secondoftwo}%
	\providecommand \translation [1]{[#1]}%
	\providecommand \BibitemOpen [0]{}%
	\providecommand \bibitemStop [0]{}%
	\providecommand \bibitemNoStop [0]{.\EOS\space}%
	\providecommand \EOS [0]{\spacefactor3000\relax}%
	\providecommand \BibitemShut  [1]{\csname bibitem#1\endcsname}%
	\let\auto@bib@innerbib\@empty
	\bibitem [{\citenamefont {Grover}(1996)}]{grover_fast_1996}%
	\BibitemOpen
	\bibfield  {author} {\bibinfo {author} {\bibfnamefont {L.~K.}\ \bibnamefont
			{Grover}},\ }\bibfield  {title} {\bibinfo {title} {A fast quantum mechanical
			algorithm for database search},\ }in\ \href
	{https://doi.org/10.1145/237814.237866} {\emph {\bibinfo {booktitle}
			{Proceedings of the {Annual} {ACM} {Symposium} on {Theory} of
				{Computing}}}},\ Vol.\ \bibinfo {volume} {Part F129452}\ (\bibinfo
	{publisher} {Association for Computing Machinery},\ \bibinfo {address} {New
		York, New York, USA},\ \bibinfo {year} {1996})\ pp.\ \bibinfo {pages}
	{212--219},\ \bibinfo {note} {iSSN: 07378017}\BibitemShut {NoStop}%
	\bibitem [{\citenamefont {Boyer}\ \emph {et~al.}(1998)\citenamefont {Boyer},
		\citenamefont {Brassard}, \citenamefont {Hoeyer},\ and\ \citenamefont
		{Tapp}}]{boyer_tight_1998}%
	\BibitemOpen
	\bibfield  {author} {\bibinfo {author} {\bibfnamefont {M.}~\bibnamefont
			{Boyer}}, \bibinfo {author} {\bibfnamefont {G.}~\bibnamefont {Brassard}},
		\bibinfo {author} {\bibfnamefont {P.}~\bibnamefont {Hoeyer}},\ and\ \bibinfo
		{author} {\bibfnamefont {A.}~\bibnamefont {Tapp}},\ }\bibfield  {title}
	{\bibinfo {title} {Tight bounds on quantum searching},\ }\href
	{https://doi.org/10.1002/(SICI)1521-3978(199806)46:4/5<493::AID-PROP493>3.0.CO;2-P}
	{\bibfield  {journal} {\bibinfo  {journal} {Fortschritte der Physik}\
		}\textbf {\bibinfo {volume} {46}},\ \bibinfo {pages} {493} (\bibinfo {year}
		{1998})},\ \bibinfo {note} {arXiv:quant-ph/9605034}\BibitemShut {NoStop}%
	\bibitem [{\citenamefont {Farhi}\ and\ \citenamefont
		{Gutmann}(1998)}]{Farhi1998}%
	\BibitemOpen
	\bibfield  {author} {\bibinfo {author} {\bibfnamefont {E.}~\bibnamefont
			{Farhi}}\ and\ \bibinfo {author} {\bibfnamefont {S.}~\bibnamefont
			{Gutmann}},\ }\bibfield  {title} {\bibinfo {title} {Analog analogue of a
			digital quantum computation},\ }\href
	{https://doi.org/10.1103/PhysRevA.57.2403} {\bibfield  {journal} {\bibinfo
			{journal} {Physical Review A - Atomic, Molecular, and Optical Physics}\
		}\textbf {\bibinfo {volume} {57}},\ \bibinfo {pages} {2403} (\bibinfo {year}
		{1998})},\ \bibinfo {note} {publisher: American Physical Society}\BibitemShut
	{NoStop}%
	\bibitem [{\citenamefont {Childs}\ and\ \citenamefont
		{Goldstone}(2004{\natexlab{a}})}]{Childs2004}%
	\BibitemOpen
	\bibfield  {author} {\bibinfo {author} {\bibfnamefont {A.~M.}\ \bibnamefont
			{Childs}}\ and\ \bibinfo {author} {\bibfnamefont {J.}~\bibnamefont
			{Goldstone}},\ }\bibfield  {title} {\bibinfo {title} {Spatial search by
			quantum walk},\ }\href {https://doi.org/10.1103/PhysRevA.70.022314}
	{\bibfield  {journal} {\bibinfo  {journal} {Physical Review A - Atomic,
				Molecular, and Optical Physics}\ }\textbf {\bibinfo {volume} {70}},\ \bibinfo
		{pages} {022314} (\bibinfo {year} {2004}{\natexlab{a}})},\ \bibinfo {note}
	{publisher: American Physical Society}\BibitemShut {NoStop}%
	\bibitem [{\citenamefont {Childs}\ and\ \citenamefont
		{Goldstone}(2004{\natexlab{b}})}]{childs_spatial_2004}%
	\BibitemOpen
	\bibfield  {author} {\bibinfo {author} {\bibfnamefont {A.~M.}\ \bibnamefont
			{Childs}}\ and\ \bibinfo {author} {\bibfnamefont {J.}~\bibnamefont
			{Goldstone}},\ }\bibfield  {title} {\bibinfo {title} {Spatial search by
			quantum walk},\ }\href {https://doi.org/10.1103/PhysRevA.70.022314}
	{\bibfield  {journal} {\bibinfo  {journal} {Physical Review A}\ }\textbf
		{\bibinfo {volume} {70}},\ \bibinfo {pages} {022314} (\bibinfo {year}
		{2004}{\natexlab{b}})},\ \bibinfo {note} {publisher: American Physical
		Society}\BibitemShut {NoStop}%
	\bibitem [{\citenamefont {Chakraborty}\ \emph {et~al.}(2020)\citenamefont
		{Chakraborty}, \citenamefont {Novo},\ and\ \citenamefont
		{Roland}}]{chakraborty_optimality_2020}%
	\BibitemOpen
	\bibfield  {author} {\bibinfo {author} {\bibfnamefont {S.}~\bibnamefont
			{Chakraborty}}, \bibinfo {author} {\bibfnamefont {L.}~\bibnamefont {Novo}},\
		and\ \bibinfo {author} {\bibfnamefont {J.}~\bibnamefont {Roland}},\
	}\bibfield  {title} {\bibinfo {title} {Optimality of spatial search via
			continuous-time quantum walks},\ }\href
	{https://doi.org/10.1103/PhysRevA.102.032214} {\bibfield  {journal} {\bibinfo
			{journal} {Physical Review A}\ }\textbf {\bibinfo {volume} {102}},\ \bibinfo
		{pages} {032214} (\bibinfo {year} {2020})},\ \bibinfo {note} {publisher:
		American Physical Society}\BibitemShut {NoStop}%
	\bibitem [{\citenamefont {Novo}\ \emph {et~al.}(2015)\citenamefont {Novo},
		\citenamefont {Chakraborty}, \citenamefont {Mohseni}, \citenamefont {Neven},\
		and\ \citenamefont {Omar}}]{novo_systematic_2015}%
	\BibitemOpen
	\bibfield  {author} {\bibinfo {author} {\bibfnamefont {L.}~\bibnamefont
			{Novo}}, \bibinfo {author} {\bibfnamefont {S.}~\bibnamefont {Chakraborty}},
		\bibinfo {author} {\bibfnamefont {M.}~\bibnamefont {Mohseni}}, \bibinfo
		{author} {\bibfnamefont {H.}~\bibnamefont {Neven}},\ and\ \bibinfo {author}
		{\bibfnamefont {Y.}~\bibnamefont {Omar}},\ }\bibfield  {title} {\bibinfo
		{title} {Systematic {Dimensionality} {Reduction} for {Quantum} {Walks}:
			{Optimal} {Spatial} {Search} and {Transport} on {Non}-{Regular} {Graphs}},\
	}\href {https://doi.org/10.1038/srep13304} {\bibfield  {journal} {\bibinfo
			{journal} {Scientific Reports 2015 5:1}\ }\textbf {\bibinfo {volume} {5}},\
		\bibinfo {pages} {1} (\bibinfo {year} {2015})},\ \bibinfo {note} {publisher:
		Nature Publishing Group}\BibitemShut {NoStop}%
	\bibitem [{\citenamefont {Chakraborty}\ \emph {et~al.}(2016)\citenamefont
		{Chakraborty}, \citenamefont {Novo}, \citenamefont {Ambainis},\ and\
		\citenamefont {Omar}}]{chakraborty_spatial_2016}%
	\BibitemOpen
	\bibfield  {author} {\bibinfo {author} {\bibfnamefont {S.}~\bibnamefont
			{Chakraborty}}, \bibinfo {author} {\bibfnamefont {L.}~\bibnamefont {Novo}},
		\bibinfo {author} {\bibfnamefont {A.}~\bibnamefont {Ambainis}},\ and\
		\bibinfo {author} {\bibfnamefont {Y.}~\bibnamefont {Omar}},\ }\bibfield
	{title} {\bibinfo {title} {Spatial {Search} by {Quantum} {Walk} is {Optimal}
			for {Almost} all {Graphs}},\ }\href
	{https://doi.org/10.1103/PhysRevLett.116.100501} {\bibfield  {journal}
		{\bibinfo  {journal} {Physical Review Letters}\ }\textbf {\bibinfo {volume}
			{116}},\ \bibinfo {pages} {100501} (\bibinfo {year} {2016})},\ \bibinfo
	{note} {publisher: American Physical Society}\BibitemShut {NoStop}%
	\bibitem [{\citenamefont {Chakraborty}\ \emph {et~al.}(2017)\citenamefont
		{Chakraborty}, \citenamefont {Novo}, \citenamefont {Di~Giorgio},\ and\
		\citenamefont {Omar}}]{chakraborty_optimal_2017}%
	\BibitemOpen
	\bibfield  {author} {\bibinfo {author} {\bibfnamefont {S.}~\bibnamefont
			{Chakraborty}}, \bibinfo {author} {\bibfnamefont {L.}~\bibnamefont {Novo}},
		\bibinfo {author} {\bibfnamefont {S.}~\bibnamefont {Di~Giorgio}},\ and\
		\bibinfo {author} {\bibfnamefont {Y.}~\bibnamefont {Omar}},\ }\bibfield
	{title} {\bibinfo {title} {Optimal {Quantum} {Spatial} {Search} on {Random}
			{Temporal} {Networks}},\ }\href
	{https://doi.org/10.1103/PhysRevLett.119.220503} {\bibfield  {journal}
		{\bibinfo  {journal} {Physical Review Letters}\ }\textbf {\bibinfo {volume}
			{119}},\ \bibinfo {pages} {220503} (\bibinfo {year} {2017})},\ \bibinfo
	{note} {publisher: American Physical Society}\BibitemShut {NoStop}%
	\bibitem [{\citenamefont {Novo}\ \emph {et~al.}(2018)\citenamefont {Novo},
		\citenamefont {Chakraborty}, \citenamefont {Mohseni},\ and\ \citenamefont
		{Omar}}]{novo_environment-assisted_2018}%
	\BibitemOpen
	\bibfield  {author} {\bibinfo {author} {\bibfnamefont {L.}~\bibnamefont
			{Novo}}, \bibinfo {author} {\bibfnamefont {S.}~\bibnamefont {Chakraborty}},
		\bibinfo {author} {\bibfnamefont {M.}~\bibnamefont {Mohseni}},\ and\ \bibinfo
		{author} {\bibfnamefont {Y.}~\bibnamefont {Omar}},\ }\bibfield  {title}
	{\bibinfo {title} {Environment-assisted analog quantum search},\ }\href
	{https://doi.org/10.1103/PhysRevA.98.022316} {\bibfield  {journal} {\bibinfo
			{journal} {Physical Review A}\ }\textbf {\bibinfo {volume} {98}},\ \bibinfo
		{pages} {022316} (\bibinfo {year} {2018})},\ \bibinfo {note} {publisher:
		American Physical Society}\BibitemShut {NoStop}%
	\bibitem [{\citenamefont {Wong}\ \emph {et~al.}(2018)\citenamefont {Wong},
		\citenamefont {Wünscher}, \citenamefont {Lockhart},\ and\ \citenamefont
		{Severini}}]{wong_quantum_2018}%
	\BibitemOpen
	\bibfield  {author} {\bibinfo {author} {\bibfnamefont {T.~G.}\ \bibnamefont
			{Wong}}, \bibinfo {author} {\bibfnamefont {K.}~\bibnamefont {Wünscher}},
		\bibinfo {author} {\bibfnamefont {J.}~\bibnamefont {Lockhart}},\ and\
		\bibinfo {author} {\bibfnamefont {S.}~\bibnamefont {Severini}},\ }\bibfield
	{title} {\bibinfo {title} {Quantum walk search on {Kronecker} graphs},\
	}\href {https://doi.org/10.1103/PhysRevA.98.012338} {\bibfield  {journal}
		{\bibinfo  {journal} {Physical Review A}\ }\textbf {\bibinfo {volume} {98}},\
		\bibinfo {pages} {012338} (\bibinfo {year} {2018})},\ \bibinfo {note}
	{publisher: American Physical Society}\BibitemShut {NoStop}%
	\bibitem [{\citenamefont {Osada}\ \emph {et~al.}(2020)\citenamefont {Osada},
		\citenamefont {Coutinho}, \citenamefont {Omar}, \citenamefont {Sanaka},
		\citenamefont {Munro},\ and\ \citenamefont
		{Nemoto}}]{osada_continuous-time_2020}%
	\BibitemOpen
	\bibfield  {author} {\bibinfo {author} {\bibfnamefont {T.}~\bibnamefont
			{Osada}}, \bibinfo {author} {\bibfnamefont {B.}~\bibnamefont {Coutinho}},
		\bibinfo {author} {\bibfnamefont {Y.}~\bibnamefont {Omar}}, \bibinfo {author}
		{\bibfnamefont {K.}~\bibnamefont {Sanaka}}, \bibinfo {author} {\bibfnamefont
			{W.~J.}\ \bibnamefont {Munro}},\ and\ \bibinfo {author} {\bibfnamefont
			{K.}~\bibnamefont {Nemoto}},\ }\bibfield  {title} {\bibinfo {title}
		{Continuous-time quantum-walk spatial search on the {Bollobás} scale-free
			network},\ }\href {https://doi.org/10.1103/PhysRevA.101.022310} {\bibfield
		{journal} {\bibinfo  {journal} {Physical Review A}\ }\textbf {\bibinfo
			{volume} {101}},\ \bibinfo {pages} {022310} (\bibinfo {year} {2020})},\
	\bibinfo {note} {publisher: American Physical Society}\BibitemShut {NoStop}%
	\bibitem [{\citenamefont {Sato}\ \emph {et~al.}(2020)\citenamefont {Sato},
		\citenamefont {Nikuni},\ and\ \citenamefont {Watabe}}]{sato_scaling_2020}%
	\BibitemOpen
	\bibfield  {author} {\bibinfo {author} {\bibfnamefont {R.}~\bibnamefont
			{Sato}}, \bibinfo {author} {\bibfnamefont {T.}~\bibnamefont {Nikuni}},\ and\
		\bibinfo {author} {\bibfnamefont {S.}~\bibnamefont {Watabe}},\ }\bibfield
	{title} {\bibinfo {title} {Scaling hypothesis of a spatial search on fractal
			lattices using a quantum walk},\ }\href
	{https://doi.org/10.1103/PhysRevA.101.022312} {\bibfield  {journal} {\bibinfo
			{journal} {Physical Review A}\ }\textbf {\bibinfo {volume} {101}},\ \bibinfo
		{pages} {022312} (\bibinfo {year} {2020})},\ \bibinfo {note} {publisher:
		American Physical Society}\BibitemShut {NoStop}%
	\bibitem [{\citenamefont {Malmi}\ \emph {et~al.}(2022)\citenamefont {Malmi},
		\citenamefont {Rossi}, \citenamefont {García-Pérez},\ and\ \citenamefont
		{Maniscalco}}]{malmi_spatial_2022}%
	\BibitemOpen
	\bibfield  {author} {\bibinfo {author} {\bibfnamefont {J.}~\bibnamefont
			{Malmi}}, \bibinfo {author} {\bibfnamefont {M.~A.~C.}\ \bibnamefont {Rossi}},
		\bibinfo {author} {\bibfnamefont {G.}~\bibnamefont {García-Pérez}},\ and\
		\bibinfo {author} {\bibfnamefont {S.}~\bibnamefont {Maniscalco}},\ }\bibfield
	{title} {\bibinfo {title} {Spatial search by continuous-time quantum walks
			on renormalized {Internet} networks},\ }\href
	{http://arxiv.org/abs/2205.02137} {\bibfield  {journal} {\bibinfo  {journal}
			{arXiv:2205.02137 [physics, physics:quant-ph]}\ } (\bibinfo {year} {2022})},\
	\bibinfo {note} {arXiv: 2205.02137}\BibitemShut {NoStop}%
	\bibitem [{\citenamefont {Lewis}\ \emph {et~al.}(2021)\citenamefont {Lewis},
		\citenamefont {Benhemou}, \citenamefont {Feinstein}, \citenamefont {Banchi},\
		and\ \citenamefont {Bose}}]{lewis_optimal_2021}%
	\BibitemOpen
	\bibfield  {author} {\bibinfo {author} {\bibfnamefont {D.}~\bibnamefont
			{Lewis}}, \bibinfo {author} {\bibfnamefont {A.}~\bibnamefont {Benhemou}},
		\bibinfo {author} {\bibfnamefont {N.}~\bibnamefont {Feinstein}}, \bibinfo
		{author} {\bibfnamefont {L.}~\bibnamefont {Banchi}},\ and\ \bibinfo {author}
		{\bibfnamefont {S.}~\bibnamefont {Bose}},\ }\bibfield  {title} {\bibinfo
		{title} {Optimal {Quantum} {Spatial} {Search} with {One}-{Dimensional}
			{Long}-{Range} {Interactions}},\ }\href
	{https://doi.org/10.1103/PhysRevLett.126.240502} {\bibfield  {journal}
		{\bibinfo  {journal} {Physical Review Letters}\ }\textbf {\bibinfo {volume}
			{126}},\ \bibinfo {pages} {240502} (\bibinfo {year} {2021})},\ \bibinfo
	{note} {publisher: American Physical Society}\BibitemShut {NoStop}%
	\bibitem [{\citenamefont {Lewis}\ \emph {et~al.}(2023)\citenamefont {Lewis},
		\citenamefont {Banchi}, \citenamefont {Teoh}, \citenamefont {Islam},\ and\
		\citenamefont {Bose}}]{lewis_ion_2023}%
	\BibitemOpen
	\bibfield  {author} {\bibinfo {author} {\bibfnamefont {D.}~\bibnamefont
			{Lewis}}, \bibinfo {author} {\bibfnamefont {L.}~\bibnamefont {Banchi}},
		\bibinfo {author} {\bibfnamefont {Y.~H.}\ \bibnamefont {Teoh}}, \bibinfo
		{author} {\bibfnamefont {R.}~\bibnamefont {Islam}},\ and\ \bibinfo {author}
		{\bibfnamefont {S.}~\bibnamefont {Bose}},\ }\bibfield  {title}
	{{\selectlanguage {en}\bibinfo {title} {Ion trap long-range {XY} model for
				quantum state transfer and optimal spatial search}},\ }\href
	{https://doi.org/10.1088/2058-9565/acd953} {\bibfield  {journal} {\bibinfo
			{journal} {Quantum Science and Technology}\ }\textbf {\bibinfo {volume}
			{8}},\ \bibinfo {pages} {035025} (\bibinfo {year} {2023})},\ \bibinfo {note}
	{publisher: IOP Publishing}\BibitemShut {NoStop}%
	\bibitem [{\citenamefont {Banchi}\ \emph {et~al.}(2020)\citenamefont {Banchi},
		\citenamefont {Zhuang},\ and\ \citenamefont {Pirandola}}]{banchi2020quantum}%
	\BibitemOpen
	\bibfield  {author} {\bibinfo {author} {\bibfnamefont {L.}~\bibnamefont
			{Banchi}}, \bibinfo {author} {\bibfnamefont {Q.}~\bibnamefont {Zhuang}},\
		and\ \bibinfo {author} {\bibfnamefont {S.}~\bibnamefont {Pirandola}},\
	}\bibfield  {title} {\bibinfo {title} {Quantum-enhanced barcode decoding and
			pattern recognition},\ }\href@noop {} {\bibfield  {journal} {\bibinfo
			{journal} {Physical Review Applied}\ }\textbf {\bibinfo {volume} {14}},\
		\bibinfo {pages} {064026} (\bibinfo {year} {2020})}\BibitemShut {NoStop}%
	\bibitem [{\citenamefont {Albanese}\ \emph {et~al.}(2004)\citenamefont
		{Albanese}, \citenamefont {Christandl}, \citenamefont {Datta},\ and\
		\citenamefont {Ekert}}]{albanese2004mirror}%
	\BibitemOpen
	\bibfield  {author} {\bibinfo {author} {\bibfnamefont {C.}~\bibnamefont
			{Albanese}}, \bibinfo {author} {\bibfnamefont {M.}~\bibnamefont
			{Christandl}}, \bibinfo {author} {\bibfnamefont {N.}~\bibnamefont {Datta}},\
		and\ \bibinfo {author} {\bibfnamefont {A.}~\bibnamefont {Ekert}},\ }\bibfield
	{title} {\bibinfo {title} {Mirror inversion of quantum states in linear
			registers},\ }\href@noop {} {\bibfield  {journal} {\bibinfo  {journal}
			{Physical review letters}\ }\textbf {\bibinfo {volume} {93}},\ \bibinfo
		{pages} {230502} (\bibinfo {year} {2004})}\BibitemShut {NoStop}%
	\bibitem [{\citenamefont {Teoh}\ \emph {et~al.}(2020)\citenamefont {Teoh},
		\citenamefont {Drygala}, \citenamefont {Melko},\ and\ \citenamefont
		{Islam}}]{teoh_machine_2020}%
	\BibitemOpen
	\bibfield  {author} {\bibinfo {author} {\bibfnamefont {Y.~H.}\ \bibnamefont
			{Teoh}}, \bibinfo {author} {\bibfnamefont {M.}~\bibnamefont {Drygala}},
		\bibinfo {author} {\bibfnamefont {R.~G.}\ \bibnamefont {Melko}},\ and\
		\bibinfo {author} {\bibfnamefont {R.}~\bibnamefont {Islam}},\ }\bibfield
	{title} {{\selectlanguage {en}\bibinfo {title} {Machine learning design of a
				trapped-ion quantum spin simulator}},\ }\href
	{https://doi.org/10.1088/2058-9565/ab657a} {\bibfield  {journal} {\bibinfo
			{journal} {Quantum Science and Technology}\ }\textbf {\bibinfo {volume}
			{5}},\ \bibinfo {pages} {024001} (\bibinfo {year} {2020})},\ \bibinfo {note}
	{publisher: IOP Publishing}\BibitemShut {NoStop}%
	\bibitem [{\citenamefont {Friis}\ \emph {et~al.}(2018)\citenamefont {Friis},
		\citenamefont {Marty}, \citenamefont {Maier}, \citenamefont {Hempel},
		\citenamefont {Holzäpfel}, \citenamefont {Jurcevic}, \citenamefont {Plenio},
		\citenamefont {Huber}, \citenamefont {Roos}, \citenamefont {Blatt},\ and\
		\citenamefont {Lanyon}}]{friis_observation_2018}%
	\BibitemOpen
	\bibfield  {author} {\bibinfo {author} {\bibfnamefont {N.}~\bibnamefont
			{Friis}}, \bibinfo {author} {\bibfnamefont {O.}~\bibnamefont {Marty}},
		\bibinfo {author} {\bibfnamefont {C.}~\bibnamefont {Maier}}, \bibinfo
		{author} {\bibfnamefont {C.}~\bibnamefont {Hempel}}, \bibinfo {author}
		{\bibfnamefont {M.}~\bibnamefont {Holzäpfel}}, \bibinfo {author}
		{\bibfnamefont {P.}~\bibnamefont {Jurcevic}}, \bibinfo {author}
		{\bibfnamefont {M.~B.}\ \bibnamefont {Plenio}}, \bibinfo {author}
		{\bibfnamefont {M.}~\bibnamefont {Huber}}, \bibinfo {author} {\bibfnamefont
			{C.}~\bibnamefont {Roos}}, \bibinfo {author} {\bibfnamefont {R.}~\bibnamefont
			{Blatt}},\ and\ \bibinfo {author} {\bibfnamefont {B.}~\bibnamefont
			{Lanyon}},\ }\bibfield  {title} {\bibinfo {title} {Observation of {Entangled}
			{States} of a {Fully} {Controlled} 20-{Qubit} {System}},\ }\href
	{https://doi.org/10.1103/PhysRevX.8.021012} {\bibfield  {journal} {\bibinfo
			{journal} {Physical Review X}\ }\textbf {\bibinfo {volume} {8}},\ \bibinfo
		{pages} {021012} (\bibinfo {year} {2018})},\ \bibinfo {note} {publisher:
		American Physical Society}\BibitemShut {NoStop}%
	\bibitem [{\citenamefont {Monroe}\ \emph {et~al.}(2021)\citenamefont {Monroe},
		\citenamefont {Campbell}, \citenamefont {Duan}, \citenamefont {Gong},
		\citenamefont {Gorshkov}, \citenamefont {Hess}, \citenamefont {Islam},
		\citenamefont {Kim}, \citenamefont {Linke}, \citenamefont {Pagano},
		\citenamefont {Richerme}, \citenamefont {Senko},\ and\ \citenamefont
		{Yao}}]{monroe_programmable_2021}%
	\BibitemOpen
	\bibfield  {author} {\bibinfo {author} {\bibfnamefont {C.}~\bibnamefont
			{Monroe}}, \bibinfo {author} {\bibfnamefont {W.}~\bibnamefont {Campbell}},
		\bibinfo {author} {\bibfnamefont {L.-M.}\ \bibnamefont {Duan}}, \bibinfo
		{author} {\bibfnamefont {Z.-X.}\ \bibnamefont {Gong}}, \bibinfo {author}
		{\bibfnamefont {A.}~\bibnamefont {Gorshkov}}, \bibinfo {author}
		{\bibfnamefont {P.}~\bibnamefont {Hess}}, \bibinfo {author} {\bibfnamefont
			{R.}~\bibnamefont {Islam}}, \bibinfo {author} {\bibfnamefont
			{K.}~\bibnamefont {Kim}}, \bibinfo {author} {\bibfnamefont {N.}~\bibnamefont
			{Linke}}, \bibinfo {author} {\bibfnamefont {G.}~\bibnamefont {Pagano}},
		\bibinfo {author} {\bibfnamefont {P.}~\bibnamefont {Richerme}}, \bibinfo
		{author} {\bibfnamefont {C.}~\bibnamefont {Senko}},\ and\ \bibinfo {author}
		{\bibfnamefont {N.}~\bibnamefont {Yao}},\ }\bibfield  {title} {\bibinfo
		{title} {Programmable quantum simulations of spin systems with trapped
			ions},\ }\href {https://doi.org/10.1103/RevModPhys.93.025001} {\bibfield
		{journal} {\bibinfo  {journal} {Reviews of Modern Physics}\ }\textbf
		{\bibinfo {volume} {93}},\ \bibinfo {pages} {025001} (\bibinfo {year}
		{2021})},\ \bibinfo {note} {publisher: American Physical Society}\BibitemShut
	{NoStop}%
	\bibitem [{\citenamefont {Tran}\ \emph {et~al.}(2020)\citenamefont {Tran},
		\citenamefont {Chen}, \citenamefont {Ehrenberg}, \citenamefont {Guo},
		\citenamefont {Deshpande}, \citenamefont {Hong}, \citenamefont {Gong},
		\citenamefont {Gorshkov},\ and\ \citenamefont {Lucas}}]{tran_hierarchy_2020}%
	\BibitemOpen
	\bibfield  {author} {\bibinfo {author} {\bibfnamefont {M.~C.}\ \bibnamefont
			{Tran}}, \bibinfo {author} {\bibfnamefont {C.-F.}\ \bibnamefont {Chen}},
		\bibinfo {author} {\bibfnamefont {A.}~\bibnamefont {Ehrenberg}}, \bibinfo
		{author} {\bibfnamefont {A.~Y.}\ \bibnamefont {Guo}}, \bibinfo {author}
		{\bibfnamefont {A.}~\bibnamefont {Deshpande}}, \bibinfo {author}
		{\bibfnamefont {Y.}~\bibnamefont {Hong}}, \bibinfo {author} {\bibfnamefont
			{Z.-X.}\ \bibnamefont {Gong}}, \bibinfo {author} {\bibfnamefont {A.~V.}\
			\bibnamefont {Gorshkov}},\ and\ \bibinfo {author} {\bibfnamefont
			{A.}~\bibnamefont {Lucas}},\ }\bibfield  {title} {\bibinfo {title} {Hierarchy
			of linear light cones with long-range interactions},\ }\bibfield  {journal}
	{\bibinfo  {journal} {Physical Review X}\ }\textbf {\bibinfo {volume} {10}},\
	\href {https://doi.org/10.1103/PhysRevX.10.031009}
	{10.1103/PhysRevX.10.031009} (\bibinfo {year} {2020}),\ \bibinfo {note}
	{publisher: American Physical Society}\BibitemShut {NoStop}%
	\bibitem [{\citenamefont {Ferrante}\ and\ \citenamefont
		{Saltalamacchia}(2014)}]{ferrante_coupon_2014}%
	\BibitemOpen
	\bibfield  {author} {\bibinfo {author} {\bibfnamefont {M.}~\bibnamefont
			{Ferrante}}\ and\ \bibinfo {author} {\bibfnamefont {M.}~\bibnamefont
			{Saltalamacchia}},\ }\bibfield  {title} {\bibinfo {title} {The {Coupon}
			{Collector}’s {Problem}}\ }(\bibinfo {year} {2014})\BibitemShut {NoStop}%
	\bibitem [{\citenamefont {Hunziker}\ and\ \citenamefont
		{Meyer}(2002)}]{hunziker_quantum_2002}%
	\BibitemOpen
	\bibfield  {author} {\bibinfo {author} {\bibfnamefont {M.}~\bibnamefont
			{Hunziker}}\ and\ \bibinfo {author} {\bibfnamefont {D.~A.}\ \bibnamefont
			{Meyer}},\ }\bibfield  {title} {{\selectlanguage {en}\bibinfo {title}
			{Quantum {Algorithms} for {Highly} {Structured} {Search} {Problems}}},\
	}\href {https://doi.org/10.1023/A:1019868924061} {\bibfield  {journal}
		{\bibinfo  {journal} {Quantum Information Processing}\ }\textbf {\bibinfo
			{volume} {1}},\ \bibinfo {pages} {145} (\bibinfo {year} {2002})}\BibitemShut
	{NoStop}%
	\bibitem [{\citenamefont {Shukla}\ and\ \citenamefont
		{Vedula}(2024)}]{shukla_efficient_2024}%
	\BibitemOpen
	\bibfield  {author} {\bibinfo {author} {\bibfnamefont {A.}~\bibnamefont
			{Shukla}}\ and\ \bibinfo {author} {\bibfnamefont {P.}~\bibnamefont
			{Vedula}},\ }\bibfield  {title} {{\selectlanguage {en}\bibinfo {title} {An
				efficient quantum algorithm for preparation of uniform quantum superposition
				states}},\ }\href {https://doi.org/10.1007/s11128-024-04258-4} {\bibfield
		{journal} {\bibinfo  {journal} {Quantum Information Processing}\ }\textbf
		{\bibinfo {volume} {23}},\ \bibinfo {pages} {38} (\bibinfo {year}
		{2024})}\BibitemShut {NoStop}%
	\bibitem [{\citenamefont {Sakurai}\ and\ \citenamefont
		{Napolitano}(2020)}]{sakurai2020modern}%
	\BibitemOpen
	\bibfield  {author} {\bibinfo {author} {\bibfnamefont {J.}~\bibnamefont
			{Sakurai}}\ and\ \bibinfo {author} {\bibfnamefont {J.}~\bibnamefont
			{Napolitano}},\ }\href@noop {} {\emph {\bibinfo {title} {Modern Quantum
				Mechanics}}}\ (\bibinfo  {publisher} {Cambridge University Press},\ \bibinfo
	{year} {2020})\BibitemShut {NoStop}%
\end{thebibliography}
\end{document}